\documentclass[12pt]{iopart}

\usepackage{amsfonts,dsfont}
\usepackage{iopams}
\usepackage{graphicx}

\newcommand{\R}{\mathbb{R}}
\newcommand{\rpn}{$\R \hbox{P}^{n-1}$}
\newcommand{\beq}{\begin{equation}}
\newcommand{\eeq}{\end{equation}}
\newcommand{\kp}{$k$-planes\ }
\newcommand{\gr}{\, $Gr(k,n)\ $}

\begin{document}

\title{Quantum Tomography twenty years later}

\date{}

\author{M. Asorey$^a$, A. Ibort$^b$, G. Marmo$^{c}$, F. Ventriglia$^c$}

\address{$^a$ Departamento de F\'{\i}sica Te\'orica,
Universidad de Zaragoza.  50009 Zaragoza, Spain.}

\address{$^b$ ICMAT \& Departamento de Matem\'{a}ticas, Universidad.
Carlos III de Madrid, Avda. de la Universidad 30, 28911 Legan\'{e}s,
Madrid, Spain }

\address{$^c$ Dipartimento di Fisica dell'
Universit\`{a} Federico II e Sezione INFN di Napoli,
Complesso Universitario di Monte S. Angelo, via
Cintia, 80126 Naples, Italy}

\eads{\mailto{asorey@unizar.es}, \mailto{albertoi@math.uc3m.es}, \mailto{marmo@na.infn.it},  
\mailto{ventriglia@na.infn.it}}


\begin{abstract}  A sample of some relevant developments that have taken place during the last twenty years in classical and quantum tomography are displayed.   We will present a general conceptual framework that provides a simple unifying mathematical picture for all of them and,  as an effective use of it, 
three subjects have been chosen that offer a wide panorama of the scope of classical  and quantum tomography:  tomography along lines and submanifolds, coherent state tomography and tomography in the abstract algebraic setting of quantum systems.   
\end{abstract}



\maketitle


\section{Introduction}  Almost twenty years ago O.V. Man'ko and V.I. Ma'ko wrote their first contributions to the foundations of Quantum Tomography (\cite{D'Ariano-Mancini}, \cite{Mancini96}, \cite{Olga97}).   
We would like to join Man'ko's celebration and highlight some of the beautiful mathematical structures in Classical and Quantum Tomography uncovered since then.   In order to do that we will present first a general conceptual framework that provides a simple unifying mathematical picture for all of them and three instances that illustrate well both the widespread scope of its applications and its conceptual unifying strength.

\subsection{The scope of tomography}
Tomography plays a very important role in nowadays science because it allows to determine
structural properties of an object by means of methods which are not invasive, i.e., they leave the object
under scrutiny in an undamaged state.   Therefore, the technique may be applied in medicine, astrophysics,
geophysics, material science, physics and nano physics.

In each field of application, tomography acquires a different form, uses different procedures and techniques so that the
unifying ideology behind it is sometimes obscured and not immediately available.

In this paper we would like to show what are its abstract mathematical aspects and framework. Due to the large and wide applicability of tomography, contributors and researchers are disseminated in a large variety of fields and
therefore it is not always possible to attribute with certainty one idea or the other to a well identified scientist, however what it is possible is to recognize one pioneer in the field in Johannes Radon and his paper of 1917 \cite{Ra17}.

In mathematical terms, the problem formulated by Radon was the following:  By integrating a function $f$, say of two variables $x$ and $y$, that satisfies some regularity conditions, along all possible straight lines of a plane, one obtains
a function depending on the lines, let us say $F(l)$, where $l$ denotes a line.   The problem that Radon solved was to  
reconstruct the initial function $f$ out of $F$.  

Of course a few questions related to the previous statements are immediate:  is every function of lines, satisfying suitable regularity conditions, obtainable by this process?  In the affirmative, is the function $f$ uniquely determined by $F$ and what  is the procedure to find it?

Some generalisations of the procedure occur immediately, going from the plane to generic manifolds, replace lines with more general submanifolds.   As straight lines may be thought of as solutions of the second order equations of motion for a free  particle, replace the free particle motion with a more general one.
When lines are replaced
 by submanifolds, describe the family of appropriate submanifolds as solutions
of suitable differential equations.  If the plane is identified with a phase-space of a particle with on degree of
freedom, is it possible to extend the procedure to general phase-spaces and interpret the original function $f$ as
a distribution probability function for a statistical mechanical system?   These questions will be explored in Sect. \ref{sec:lines}

By thinking of the plane as an Abelian vector group, one can introduce the Wigner-Weyl formulation of Quantum Mechanics
and consider the plane (phase-space) as the carrier space of a quantum mechanical picture in terms
of ``quantizer-dequantizer'' description.   In this manner the original treatment of the problem devised by Radon
is carried over to the quantum situation by considering  a composition of the Radon procedure with the Wigner-Weyl map.  
The presence of Weyl's map allows to introduce also coherent states and therefore extend the various
applications to quantum optics.

When the original function $f$ is thought of as a Wigner function, then function on lines $F$ becomes a probability
distribution.  Thus  the probabilistic aspects of Quantum Mechanics enters for free in the ``tomographic picture'', therefore many of the questions naturally associated with probabilities also enter this description, most notably entropies and thermodynamical concepts.   See for instance \cite{Pedatom, VentriPositive} for a discussion of these aspects and many others as well as a proof of the completeness of the the tomographic picture of Quantum Mechanics.
 
This approach to the tomographic description of quantum mechanical systems will be discussed in Sect. \ref{sec:coherent} where it will be shown that the  coherent-state tomographic point of view provides a unifying description that encompasses various descriptions of states of quantum systems like the Wigner quasi-distribution function \cite{Wi32}, the Husimi-Kano $K-$function \cite{Husimi1940,Kano56},
and Sudarshan's $\phi$-diagonal coherent-state representation \cite{Sudarshan1963,Glauber1963}.
 
\subsection{General considerations on the mathematical background of tomography}\label{sec:general}

At this stage, and because of the widespread use of tomographic notions, it is relevant to 
extract the main structures behind the many use of it.  

In order to accomplish that
we will replace the space of functions subjected to a tomographic treatment with a (subset of a) generic vector
space $V$.  The integration procedure along a line may be considered as a linear
functional on $V$, i.e., an element $\alpha$ in the dual space $V^*$.  The family of all
`lines' identifies a subset in  $V^*$, say $\mathcal{N}$.  Now with any vector $v\in V$, it is
possible to associate a function $F_v$ on $\mathcal{N}$ by setting,
\begin{equation}\label{general_tomogram}
F_v(\alpha ) = \alpha (v) \, .
\end{equation}

The `reconstruction process' may now be formulated in the following terms.  Given a function
$F(\alpha )$ is it possible to find a unique vector $v\in V$ such that $F = F_v$?  Which functions
on $\mathcal{N}$ are associated with vectors in $V$?

When the vector space $V$ itself is realised as a space of functions on a manifold $M$, our
procedure associates a function $F$ on $\mathcal{N}$ out of a function $f$ on $M$. Thus the Radon
transform from a vector space $V = \mathcal{F}(M)$ and uses a subspace of linear functionals, immersion 
of $\mathcal{N}$ into the dual space of $\mathcal{F}(M)$, so that with any function $\varphi \in \mathcal{F}(M)$, a function
$F_\varphi \in \mathcal{F}(\mathcal{N})$ can be constructed.

The reconstruction procedure amounts to invert the previous map.  It is now clear that $M$ should
be a space with a measure, to be able to define integrals, and the measure should have regularity properties
so that it also induces measures on submanifolds of $M$.  the chosen family of submanifolds should have a manifold
structure so that we identify $\mathcal{N}$ and a measure on it.   

All various ingredients should be judiciously chosen so that the direct map from $\mathcal{F}(M)$ to $\mathcal{F}(\mathcal{N})$ should have an inverse.  it may happen, and it does, that both the domain, in $\mathcal{F}(M)$, and the codomain in $\mathcal{F}(\mathcal{N})$ should be properly restricted to achieve the invertibility of the constructed maps.

In what follows we are going to carefully explain how these domains and codomains are chosen in specific instances of classical and quantum tomography.

As tomography allows to provide a unified picture of classical and quantum systems, we shall first indulge on some general comments concerning the mathematical description of classical and quantum systems.

\subsection{Description of classical and quantum systems and tomography}

A formal description of a physical system requires the identification of the following entities:
\begin{enumerate}
\item[i.]  A space of observables $\mathcal{A}$.

\item[ii.] A space of states $\mathcal{S}$,

\item[iii.] A pairing between states and observables which gives a real number, the outcome of a measurement of an
observable $A$ of the system in the state $\rho$.
\end{enumerate}

To take into account specific aspects of the measuring process, we require that a map $\mu \colon \mathcal{A} \times \mathcal{S} \to \mathrm{Bo}(\mathbb{R})$, where $\mathrm{Bo}(\mathbb{R})$ denotes the space of probability measures on the Borelian sets of $\mathbb{R}$, is given.  Then, the probability $P_{A,\rho}(\Delta)$ that the outcome of measuring $A$ on the state $\rho$ will lie in the interval $\Delta$ is given by the measure of the interval $\Delta$ with respect to the Borelian measure $\mu(A,\rho)$.  If the measure $\mu(A,\rho)$ is absolutely continuous with respect to the standard Lebesgue measure $dx$ on $\mathbb{R}$, then we will get
$$
P_{A,\rho}(\Delta) =  \int_{\Delta} \mu_{A,\rho}(x) dx \, ,
$$
where $\mu_{A,\rho}(x) $ is the density determined by the measure $\mu(A,\rho)$, i.e., the Radon-Nikodym derivative of $\mu(A,\rho)$ with respect to $dx$.

As for the evolution of the system, we require that a flow structure is given on $\mathcal{A}$:
$$
\Phi(t,s) \colon \mathcal{A} \to \mathcal{A} \, ,
$$
such that 
$$
\Phi(t,t) = \mathbb{I} \, , \qquad \Phi(t,s)\circ \Phi(s,r) = \Phi(t,r) \, .
$$
What we have stated are minimal requirements to have a reasonable description of physical systems.

Often additional structures are imposed on observables, states and evolution, as a consequence of the experiments performed on them.   For instance, the space of observables of a quantum system can be required to carry the structure of a Jordan algebra.  With the help of ifs derivation algebra and a compatibility condition we can construct a Lie-Jordan algebra and from here a $C^*$-algebra.  

Where the previous assumptions are made, states are positive normalized elements in the dual space of $\mathcal{A}$ and they constitute a convex body.

The evolution is required to provide automorphisms of the algebra $\mathcal{A}$.  When the flow $\Phi(t,s)$ depends only on the differences $t-s$, we get a one-parameter group (or semigroup) of transformations, and from here we derive an infinitesimal generator for the flow that allows to write the evolution in terms of a differential equation.

In Sect. \ref{sec:algebras} it will be discussed the tomographic description of quantum states in such abstract algebraic setting.  A group representation will be used to obtain the tomographic representation and its relation with other tomographic pictures will be analyzed.


\section{Tomography along lines and submanifolds}\label{sec:lines}

As it was discussed in the introduction, the Radon transform as originally formulated solves the following
problem: to reconstruct a function of two variables from  its integrals over arbitrary lines.
The original Radon transform \cite{Ra17} maps functions defined 
on a two-dimensional plane onto functions defined on a two-dimensional
cylinder. The key feature is that the transform is invertible, i.e. the function can always 
be fully recovered from its Radon transform.
There exist several important generalizations of the Radon transform \cite{Gelf, mihlin, Helgason74,Helgason80, Helgason84}. More recent analysis have focused on affine
symplectic transforms \cite{Mancini95}, on the deep relationship with classical systems and classical dynamics 
\cite{Olga97, tomocylinder}, and on the study of marginals along curved submanifolds \cite{tomocurved}.

The aim of this work is to review the study of generalizations of the Radon
transform to multidimensional phase spaces and to frameworks 
based on marginals along curves or surfaces described by \emph{quadratic}
equations.  
There are interesting applications  of these generalizations to both classical and quantum systems. 
For classical systems  the Radon transform of  probability densities
in phase space of a classical particle can be used to obtain the initial probability densities.

The generalization for  quantum systems  can be achieved by means of the corresponding
tomograms of Wigner functions in the phase space \cite{quantomogram}. Notice that the major
advantage of the tomographic approach to quantum systems is that it allows a formulation of quantum
systems in terms of pure classical probability distributions in phase space. 

\subsection{Radon transform on the plane} 

\medskip

\begin{figure}[t]
\begin{center}
\includegraphics[width=9cm]{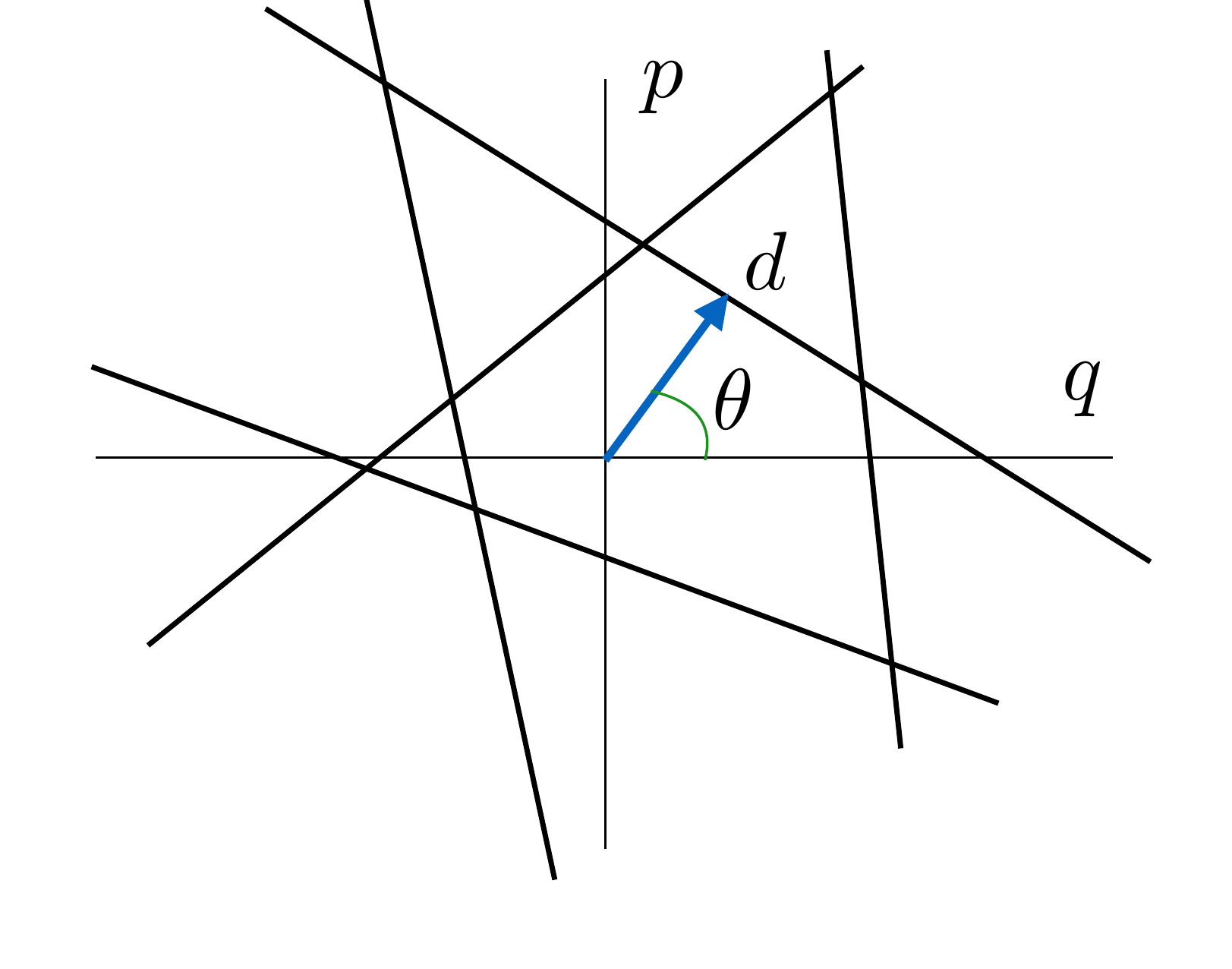}
\end{center}
\caption{Tomography on the plane. }
\label{fig:plane}
\end{figure}

In the $\mathbb{R}^2 $ plane, a line 
\begin{equation}
d = q \cos\theta +p \sin\theta, \qquad (q,p)\in \R^2
\label{line}
\end{equation}
is parametrized by the  distance  of the line to the origin $d\in \R^+$  
and  its angle $\theta$ with a reference line crossing the origin. 
Considering the phase $\e^{i \theta}\in \mathbb{S}^1$, the family of lines acquires a cylinder manifold structure
$\mathbb{R}^+\!\times\mathbb{S}^1$  (See Fig.\ \ref{fig:plane}). 
There is an alternative  group theoretical description of the manifold of straight lines
of $\R$,
which can be used for higher dimensional generalizations.
 The Euclidean group
$\mathrm{E}(2)$ acts transitively on the set of lines in the plane,
with a stability group given by the translations  $\R$ along the line 
and the reflections with respect to it
$\mathbb{Z}_2$. Therefore the family of lines is given by
$\mathrm{E}(2)/(\mathbb{Z}_2\times\mathbb{R})$, which is isomorphic to \ $\mathbb{R}^+\!\times\mathbb{S}^1$.

It is interesting to observe that depending on which subgroup we use to
quotient $E(2)$ we either obtain the plane $\R^2$ or the cylinder $\R^+\!\times\mathbb{S}^1$ 
\begin{equation}
\mathbb{R}^2 = \mathrm{E}(2)/O(2), \qquad
\mathbb{R}^+\!\times\mathbb{S}^1 = \mathrm{E}(2)/({\mathbb Z}_2\times\mathbb{R}).
\end{equation}
The Radon transform is a map of
 $L^1$  functions  of $\R^2$ into  $L^1$ functions
 on $\mathbb{R}^+\times\mathbb{S}^1$ given by

\begin{equation}
\widehat{f}(d,\theta)=\int_{-\infty}^{+\infty} ds\, f(s\sin\theta
+d\cos\theta,-s\cos\theta+ d\sin\theta) ,
\end{equation}
where $s$ is the parameter along the line given by \eref{line}.
The inversion formula, as given by Radon \cite{Ra17}, amounts to consider first
the average value of $F$ on all lines tangent to the circle of
center $(q,p)$ and radius $r$
\begin{equation}
\widetilde{{f}}_r (p,q)=\frac{1}{2\pi}\int_0^{2\pi}d\theta\, \widehat{f}(q \cos \theta + p\sin\theta +
r,\theta) 
\end{equation}
and then use the Hilbert transform
\begin{equation}
f(q,p)=-\frac{1}{\pi}\int _0^\infty \frac{dr}{r}\, {\widetilde{f}'_r(p,q)} ,
\end{equation}
to reconstruct the original function.  

Later on we will provide an alternative expression
for the inverse Radon transform in an affine language that extends easily to
more general situations, see Eq. (\ref{eq:invgen}).   
Now we will discuss how the Radon transform can be generalized to higher dimensional spaces.

\subsection{Radon transform on the Euclidean space $\R^n$}

The manifold of straight lines of $\R^n$ crossing the origin  is the real projective space \rpn. This manifold can also be identified with the space of hyperplanes of dimension $n-1$ crossing the origin,  and with the quotient of the $\mathbb{S}^{n-1}$ unit sphere by the
reflection symmetry $\mathbb{Z}_2$,
\beq
\hbox{\rpn}=\mathbb{S}^{n-1}/\mathbb{Z}_2.
\eeq
In the $\mathbb{R}^n $ space, any hyperplane of dimension $n-1$
\begin{equation}
d =| \mathbf n\cdot {\mathbf x}|, \qquad {\mathbf x} \in \R^n
\end{equation}
is parametrized by its distance to the origin $d\in \R^+$  
and  a unit normal vector $\mathbf n\in \mathbb{S}^{n-1}$ of $\R^n$, or what is equivalent a real number $\pm d=\mathbf n\cdot {\mathbf x}$ and a ray of the projective $\R$P$^{n-1}$. Thus,  the family $(n-1)$-hyperplanes of $\R^n$ acquires a  manifold structure
\beq
\R^+ \times \mathbb{S}^{n-1}\equiv \R/\mathbb{Z}_2 \times \mathbb{S}^{n-1}\equiv \R \times \mathbb{S}^{n-1}/\mathbb{Z}_2\equiv  \R\times\hbox{\rpn},
\eeq
i.e. the manifold of all $(n-1)$-hyperplanes of $\R^n$  which can be identified with  $\R\times$\rpn. 
 There is  another
interesting   group theoretical description of this manifold  
which is useful for the analysis of Radon transform. The Euclidean group
$\mathrm{E}(n)$ acts transitively on the set of $(n-1)-$hyperplanes 
with a stability group given by the Euclidean transformations along the $(n-1)$-hyperplane
and the $\mathbb{Z}_2$ reflections with respect to it. Therefore, the family of $(n-1)$-hyperplanes is given \begin{equation}
\R\times\hbox{\rpn} = \mathrm{E}(n)/(\mathbb{Z}_2\times \mathrm{E}(n-1)).
\end{equation}
It is interesting to observe that depending on which subgroup we use to
quotient we either obtain the Euclidean space $\R^n$ or the  space of its $(n-1)$-hyperplanes, 
 \begin{equation}
\mathbb{R}^n = \mathrm{E}(n)/O(n), \qquad
\R\times\hbox{\rpn} = \mathrm{E}(n)/(\mathbb{Z}_2\times \mathrm{E}(n-1)).
\end{equation}

The natural generalization of the Radon transform for higher dimensional 
Euclidean spaces $\R^n$ is a map of
 $L^1$  functions  of $\R^n$ into  $L^1$ functions
 on $\R\times$\rpn given by
\beq
\widehat{f}(\xi)= \int_{{x}\in \xi} d\mathbf{x} f(\mathbf{x})
\eeq
where the $(n-1)$-dimensional integral is over the points of $\mathbf{x}\in\R^n$ contained in the  
$(n-1)$-hyperplane $\xi=(d,{\mathbf{n}})\in\R^+\times \mathbb{S}^{n-1}$.

There is a dual Radon map which maps   $L^1$  functions  of $\R\times$\rpn into  $L^1$ functions
 on $\mathbb{R}^n$ given by
\beq
\widetilde{f}(x)= \int_{x\in \xi_x} d\mu(\xi_x) \widehat{f}(\xi_x)=\frac{\Gamma(\frac{n}{2})}{2 \pi^{\frac{n}{2}}}\int_{\mathbb{S}^{n-1}} d {\mathbf n}\, \widehat{f}({\mathbf x}\cdot {\mathbf n},{\mathbf n}),
\eeq
where the $(n-1)$-dimensional integral is extended to all $(n-1)$-hyperplanes $\xi_x$ crossing at the point $x$ with the probability measure induced from the measure of the projective space \rpn obtained by projection of
the standard homogeneous measure of $\mathbb{S}^{n-1}$.

The inverse of the generalized Radon transform can then given in terms of the dual Radon map. However, the explicit 
formulas originally given by Radon and John \cite{Ra17, John} are different for the cases of even and odd
Euclidean spaces,

\beq
f(x)=\displaystyle {\frac{{\Gamma(1/2)}}{(4\pi)^{\frac{n-1}{2}}{\Gamma(n/2)}}} 
 \left\{ 
 \begin{array}{lc}
 \displaystyle\mathbf\Delta^{\frac{n-2}{2}} \displaystyle\int_0^\infty \frac{d r}{r}  (-\partial_r){\widetilde{f}_r}(y),&  \quad \mbox{for} \quad n \quad \mbox{even}\phantom{\Bigg]} \\
  \displaystyle{\mathbf\Delta}^{\frac{n-1}{2}} \widetilde{f}(x),& \quad \mbox{for} \quad  n \quad \mbox{odd,}
\end{array}  \right.
\eeq
where $\mathbf\Delta=-\sum_{i=1}^n\partial_i\partial^i$ and  
\beq
{\widetilde{f}_r}(x)= \int_{x\in \xi_x} d\mu(\xi_x) \widehat{f}({\mathbf x}\cdot {\mathbf n}+r,{\mathbf n})=\frac{\Gamma(\frac{n}{2})}{2 \pi^{\frac{n}{2}}}\int_{\mathbb{S}^{n-1}} d {\mathbf n}\, \widehat{f}({\mathbf x}\cdot {\mathbf n}+r,{\mathbf n}).
\eeq
Both cases can be described in a unified way in terms of pseudo-differential operators by the  following formula \cite{Helgason74, Helgason80,Facchi}
\beq
f(x)= {\frac{{\Gamma(\frac12)}}{(4\pi)^{\frac{n-1}{2}}{\Gamma(\frac{n}{2})}}}   \displaystyle{\mathbf\Delta}^{\frac{n-1}{2}} \widetilde{f}(x),
\eeq
or using an equivalent formulation in terms of  Riesz potentials \cite{Ludwig}  
\beq
f(x)=\frac1{\pi^{{n-1}} }
 \frac{\Gamma(n-\frac{1}{2})}{\Gamma(\frac{n}{2})\Gamma(\frac12-\frac{n}{2})}\int_{\R^n} d^n y\, \frac1{\|\mathbf {x-y}\|^{2n-1}} \widetilde{f}(y). 
\eeq

\subsection{Radon transform on $k$-planes of  Euclidean space $\R^n$}

The manifold of hyperplanes of dimension $k$ of $\R^n$ (k-planes) crossing the origin  is the real Grassmannian manifold \gr , which
can also be identified with
\beq
\hbox{\gr}=\frac{O(n)}{O(n-k)\times O(k)}.
\eeq
The family of \kp  of $\R^n$ acquires a  manifold structure
\begin{equation}
\R^{n-k}\times\hbox{\gr} = \frac{\mathrm{E}(n)}{\mathrm{E}(k)\times O(n-k)}.
\end{equation}

The generalized  Radon transform  is a map of
 $L^1$  functions  of $\R^n$ into  $L^1$ functions
 on $\R^{n-k}\times$\gr\  given by
\beq
\widehat{f}(\xi)= \int_{{x}\in \xi} d\mathbf{x} f(\mathbf{x})
\eeq
where the $k$-dimensional integral is over the points of $\mathbf{x}\in\R^n$ contained in the  hyperplane 
line $\xi\in$\gr.

There is a dual Radon map which maps   $L^1$  functions  of $\R^{n-k}\times$\gr\  into  $L^1$ functions
 on $\mathbb{R}^n$ given by
\beq
\widetilde{f}(x)= \int_{x\in \xi_x} d\mu(\xi_x) \widehat{f}(\xi_x),
\eeq
where the $k(n-k)$-dimensional integral is extended to all $k$-hyperplanes $\xi_x$ crossing at the point $x$ with
the probability measure induced from the measure of the Grassmannian manifold \gr obtained by projection of
the Haar measure of $O(n)$.

The inverse of the generalized Radon transform can then given in terms of the dual Radon map 
 \cite{Helgason74, Helgason80,Facchi}
\beq
f(x)= {\frac{{\Gamma(\frac{n-k}{2})}}{(4\pi)^{\frac{k}{2}}{\Gamma(\frac{n}{2})}}}   \displaystyle{\mathbf\Delta}^{\frac{k}{2}} \widetilde{f}(x),
\eeq
or using an equivalent formulation in terms of  Riesz potentials \cite{Ludwig}  
\beq
f(x)=\frac1{\pi^{\frac{n+k}{2}} }
 \frac{\Gamma(\frac{n-k}{2})\Gamma(\frac{n+k}{2})}{\Gamma(\frac{n}{2})\Gamma(-\frac{k}{2})}\int_{\R^n} d^n y\, \frac1{\|\mathbf {x-y}\|^{n+k}} \widetilde{f}(y) 
\eeq
The \kp are geodesic submanifolds of $\R^n$. The generalization of Radon transform for complete geodesic manifolds 
is possible. Let us analyze the simplest cases of hyperbolic spaces $\mathbb{H}^n$.

\subsection{Radon transform in hyperbolic spaces $\mathbb{H}^n$}
The hyperbolic space $\mathbb{H}^n$ is a negative constant curvature Riemanian manifold which can be identified
with an hyperboloid inmersed in a $n+1$ Minkowski space-time. This space is geodesically complete. Let us denote
by $\Xi^\xi_x$ the space of all geodesics which are tangent to a $k$-plane $\xi$ of the tangent space at x.  For a given k-plane $\xi\subset T_x \mathbb{H}^n $
the corresponding submanifold in $\Xi^\xi_x$ is a totally geodesic manifold.  Due to the special properties of $\mathbb{H}^n$, the identification of the space $\Xi_x$ of all totally geodesic submanifolds of the same type $\Xi^\xi_x$ containing $x$ and the Grassmannian \gr is one-to-one.  The group of isometries of $\mathbb{H}^n$ is the Lorentz group $O(n,1)$ and the subgroup of isometries which leave invariant a $k$-dimensional geodesic submanifold $\Xi^\xi_x$ can be identified with $O(n-k)\times O(k,1)$.  The space of totally 
geodesic submanifolds  $\Xi^k$ can be identified with
\beq
\Xi^k=\frac{O(n,1)}{O(n-k)\times O(k,1)}.
\label{ident}
\eeq
The generalized Radon transform
can be defined as
\beq
\widehat{f}(\Xi^\xi)= \int_{{x}\in \Xi^\xi} d\mu_{\Xi^\xi}(\mathbf{x}) f(\mathbf{x})
\eeq
where the $k$-dimensional integral is over the geodesically complete submanifold $\Xi^\xi$ of $\mathbb{H}^n$ endowed with the  probability measure defined by the Riemannian metric induced by its immersion into $\mathbb{H}^n$.

In the case of hyperbolic spaces $\mathbb{H}^n$ the dual map 
\beq
\widetilde{f}(x)= \int_{\Xi^\xi_x\in  \Xi_x^k} d\mu(\Xi^\xi_x) \widehat{f}(\Xi^\xi_x),
\eeq
is defined in terms of the probability measure induced from the Haar measure of $O(n,1)$ by means of the identification
\eref{ident}.

In  hyperbolic spaces $\mathbb{H}^n$ the inverse Radon transform  is obtained by \cite{Helgason80, Helgason84}
\beq
\hskip-2.5cm f(\mathbf{x})= \frac{\Gamma(\frac{n-k}{2})}{(4\pi)^\frac{k}{2} {\Gamma(\frac{n}{2})}}
 \left\{ \begin{array}{lc}
  \displaystyle P_k(\mathbf\Delta) \widetilde{f}(x),& \mbox{for} \, k \, \mbox{even}\phantom{\Bigg]} \\
\displaystyle\frac{\Gamma(\frac{n-1}{2})}{2 \pi^{\frac{n}{2}}\Gamma(\frac{1}{2})}
\int_{\mathbb{H}^n} d \mathbf{y}  \sinh^{1-n}d(\mathbf{y},\mathbf{x}) \cosh d(\mathbf{y},\mathbf{x}) P_k(\mathbf\Delta)  \widetilde{f}(\mathbf{y})),& \mbox{for}\, k\, \mbox{odd,}
\end{array} \right.
\eeq
where $ d(\mathbf{y},\mathbf{x})$ denotes the geodesic distance between two points $ \mathbf{y},\mathbf{x}\in \mathbb{H}^n$, and 
\beq
P_k(\mathbf\Delta)= 
\left\{\begin{array}{lc} \displaystyle  \prod_{l=0}^{[\frac{k}{2}]-1} \left( (k-2l-n)(k-2l-1)+ \mathbf\Delta)\right)
&\quad k>1\\
\ \mathbf\Delta &\quad k=1,
\end{array}\right.
\eeq
 $[\frac{k}{2}]$ being the integer part of $\frac{k}{2}$.
 
A similar analysis can be applied to spaces with constant positive curvature \cite{Helgason80}.

\subsection{Affine symplectic tomography on the plane}
\label{sec-tomogplane}

The generalization of Radon transform to more general tomographic submanifolds which are not totally
geodesic it is possible by changing the perspective of Radon transform and to consider it as
a special case of Fourier transform which allow  then the use  of harmonic analysis. 

Let us define the Radon transform in the affine language
(we called it tomographic map) \cite{Ra17,Gelf}
\begin{eqnarray}
\widehat{f}(\lambda ,\mu,\nu) &=& \left\langle \delta(\lambda -\mu q-\nu
p)\right\rangle \nonumber\\
& =& \int_{\mathbb{R}^2}  dq\,dp\, f(q,p) \delta(\lambda -\mu q-\nu p)  ,
\label{eq:radondef}
\end{eqnarray}
where $\delta$ is the Dirac function and the parameters $\lambda , \mu,
\nu \in \mathbb{R}$. 

The main advantage  of this affine approach is that the 
 inverse transform acquires a simple form
 \cite{Gelf}
\begin{eqnarray}
f(q,p) &=&  \int_{\mathbb{R}^3}\frac{d\lambda\, d\mu\, d\nu}{(2\pi)^2}\, \widehat{f}(\lambda ,\mu,\nu) e^{i(\lambda -\mu
q-\nu p)}  . \quad \label{eq:invgen}
\end{eqnarray}

We remark the affine tomographic map is homogeneous
\begin{eqnarray}
\widehat{f}(s \lambda ,s \mu,s \nu) =
\frac{1}{|s|}\widehat{f}(\lambda ,\mu,\nu).
\label{eq:homog}
\end{eqnarray}

\subsection{Tomography by  hyperplanes}
\label{sec-hyperplane}

{The above construction can be  generalized for higher dimensional spaces
in a straightforward way}.
Let us consider a function $f(\mathbf{x})$ on the $n$-dimensional space $\mathbf{\mathbf x}
\in \mathbb{R}^n$. {It is possible to   reconstruct the function $f$ from its integrals
over arbitrary $(n-1)$-dimensional hyperplanes.}

A generic hyperplane is given by the equation
\begin{equation}
\lambda  -\boldsymbol{\mu}  \cdot \mathbf{x} = 0,
\label{eq:hypplan}
\end{equation}
with $\lambda \in\mathbb{R}$ and $\boldsymbol{\mu}\in\mathbb{R}^n$.

We recall that we have seen in subsection 2.3 that the space of (n-1)-hyperplanes is diffeomorphic to $\mathbb{R}\times\mathbb{R}P^{n-1}$, which is the set of pairs $(\lambda,[\boldsymbol{\mu}])$, where $\lambda\in \R$
is an arbitrary real number and $[\boldsymbol{\mu}]$
denotes the projective ray of the vector $\boldsymbol{\mu}\in \R^n$.

The Radon transform is given by
\begin{eqnarray}
\widehat{f}(\lambda ,\boldsymbol{\mu}) &=& \left\langle \delta(\lambda -\boldsymbol{\mu} \cdot \mathbf{x})\right\rangle \nonumber\\
& =& \int_{\mathbb{R}^n}  d^n\mathbf{x}\, f(\mathbf{x}) \delta(\lambda -\boldsymbol{\mu}\cdot \mathbf{x})  ,
\label{eq:radondefn}
\end{eqnarray}
where $\delta$ is the Dirac function and the parameters $\lambda \in
\mathbb{R}$ and $\boldsymbol{\mu} \in \mathbb{R}^n$. When $n=2$
Eq.~(\ref{eq:radondef}) is recovered.

It is very easy to show that the inverse transform of
(\ref{eq:radondefn}) reads
\begin{eqnarray}
f(\mathbf{x}) &=&  \int_{\mathbb{R}^{n+1}}\frac{d\lambda\, d^n \boldsymbol{\mu}}{(2\pi)^n}\,  \widehat{f}(\lambda ,\boldsymbol{\mu}) e^{i(\lambda -\boldsymbol{\mu}\,\cdot\,
\mathbf{x})}  . \quad \label{eq:invgenn}
\end{eqnarray}

In quantum mechanics this version of affine symplectic Radon transform can be applied to Wigner functions
providing a center of mass tomography \cite{archi}.  Moreover in \cite{Am09} the classical limit for this tomogram was calculated under some additional conditions.

\subsection{{Tomography with more general submanifolds}}
\label{sec-submanifolds}

{A simple  mechanism which allow non-linear generalizations of Radon transform
is the combination of the standard affine transform with a diffeomorphism
of the underlying $\mathbb{R}^n$ space.}
Let us consider a function $f(\mathbf{x})$ on the $n$-dimensional space $\mathbf{x}
\in \mathbb{R}^n$. The problem one needs to solve is to reconstruct $f$ from
 its integrals over an $n$-parameter family of  codimension one submanifolds.

\begin{figure}[t]
\begin{center}
\includegraphics[width=8cm]{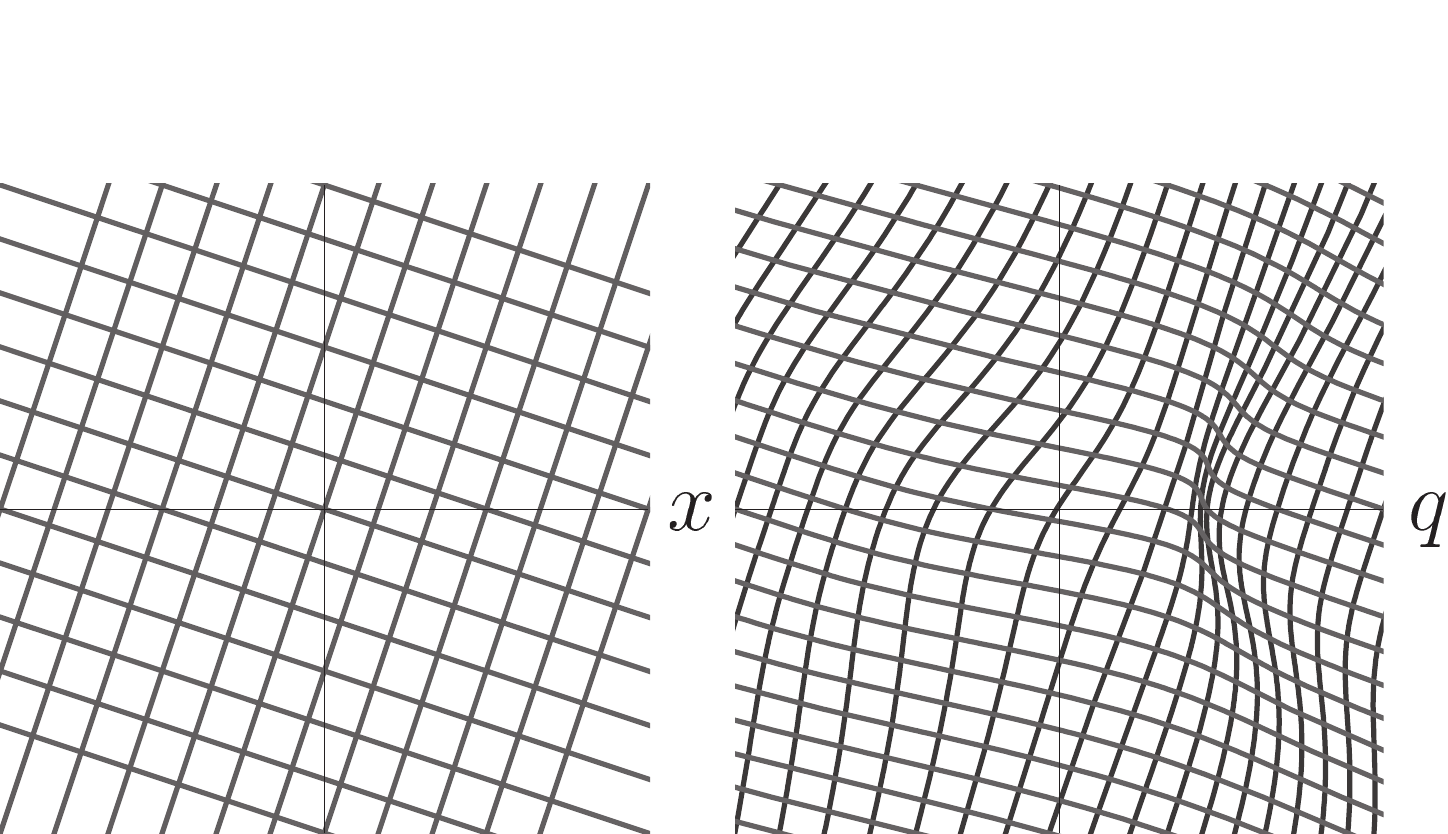}
\end{center}
\caption{Diffeomorphism $\boldsymbol{\varphi}$ on the plane $\mathbf{x}=\boldsymbol\varphi(\mathbf{q})$. }
\label{fig:diffeo}
\end{figure}

We can construct such a family by diffeomorphic deformations of
the hyperplanes of $\mathbb{R}^n$
\begin{equation}
\lambda -\boldsymbol{\mu} \cdot \mathbf{x} = 0, \label{eq:planes}
\end{equation}
with $\lambda\in\mathbb{R}$ and $\boldsymbol{\mu}\in\mathbb{R}^n$. { Let us 
consider a diffeomorphism $\boldsymbol{\varphi}$ of $\mathbb{R}^n$}
\begin{equation}
\mathbf{q}\in\mathbb{R}^n\mapsto \mathbf{x}=\boldsymbol{\varphi}(\mathbf{q})\in\mathbb{R}^n.
\end{equation}
The hyperplanes (\ref{eq:planes}) are deformed {by means of
$\boldsymbol{\varphi}$} into a family of
submanifolds in the $\mathbf{q}$ space (see Fig. \eref{fig:diffeo})
\begin{equation}
\lambda -\boldsymbol{\mu} \cdot \boldsymbol{\varphi}(\mathbf{q}) = 0.
\end{equation}

The Radon transform {can be rewritten as}
\begin{eqnarray}
\!\!\!\!\!\!\!\!\!\!\!\!\widehat{f}(\lambda,\boldsymbol{\mu}) &=& \left\langle \delta(\lambda-\boldsymbol{\mu} \cdot \boldsymbol{\varphi}(\mathbf{q}))\right\rangle \nonumber\\
&=& \int_{\mathbb{R}^n}  d^n\mathbf{x}\, f(\mathbf{x}) \delta(\lambda-\boldsymbol{\mu}\cdot \mathbf{x})  
\nonumber\\
&=& \int_{\mathbb{R}^n} d^n\mathbf{q}\,  f(\boldsymbol{\varphi}(\mathbf{q})) \delta(\lambda-\boldsymbol{\mu}\cdot
\boldsymbol{\varphi}(\mathbf{q}))J(\mathbf{q})  ,
\label{eq:radondefns0}
\end{eqnarray}
where
\begin{equation}\label{eq:Jacobian}
J(\mathbf{q})=\left|\frac{\partial x_i}{\partial
q_j}\right|=\left|\frac{\partial \varphi_i(\mathbf{q})}{\partial
q_j}\right|
\end{equation}
is the Jacobian of the transformation and $ f(x)$ is an arbitrary function.

Now, since 
\begin{equation}
 f(\mathbf{x}) d^n \mathbf{x} = f(\boldsymbol{\varphi}(\mathbf{q})) J(\mathbf{q}) d^n \mathbf{q},
\end{equation}
the Radon transform can be rewritten as
\begin{eqnarray}
\widehat{f}(\lambda,\boldsymbol{\mu}) &=& \left\langle \delta(\lambda-\boldsymbol{\mu} \cdot \boldsymbol{\varphi}(\mathbf{q}))\right\rangle \nonumber\\
& =& \int_{\mathbb{R}^n} d^n\mathbf{q} \, f(\mathbf{q}) \delta(\lambda-\boldsymbol{\mu}\cdot \boldsymbol{\varphi}(\mathbf{q})) \, ,
\label{eq:radondefns}
\end{eqnarray}
with $\lambda\in \mathbb{R}$ and $\boldsymbol{\mu} \in \mathbb{R}^n$.

The inverse transform is given by
\begin{eqnarray}
f(\mathbf{q}) &=&   
\int_{\mathbb{R}^{n+1}} \frac{d\lambda\, d^n\boldsymbol{\mu}}{(2\pi)^n}\, \widehat{f}(\lambda,\boldsymbol{\mu}) J(\mathbf{q}) e^{i(\lambda-\boldsymbol{\mu}\cdot
\boldsymbol{\varphi}(\mathbf{q}))}  , \quad
\label{eq:invgens}
\end{eqnarray}
with a modified kernel
\begin{equation}\label{eq:kernel}
\!\!\!\!\!K(\mathbf{q};\lambda ,\boldsymbol{\mu})\! =\! J(\mathbf{q}) e^{i(\lambda -\boldsymbol{\mu}\cdot \boldsymbol{\varphi}(\mathbf{q}))}=
\left|\frac{\partial \varphi_i(\mathbf{q})}{\partial q_j}\right|
e^{i(\lambda -\boldsymbol{\mu}\cdot \boldsymbol\varphi (\mathbf{q}))}.
\end{equation}

{We can now consider different applications of this deformed
generalizations of Radon transform}.

\subsection{Tomography by circles in the plane}
\label{sec-circles}

\begin{figure}[t]
\begin{center}
\includegraphics[width=8cm]{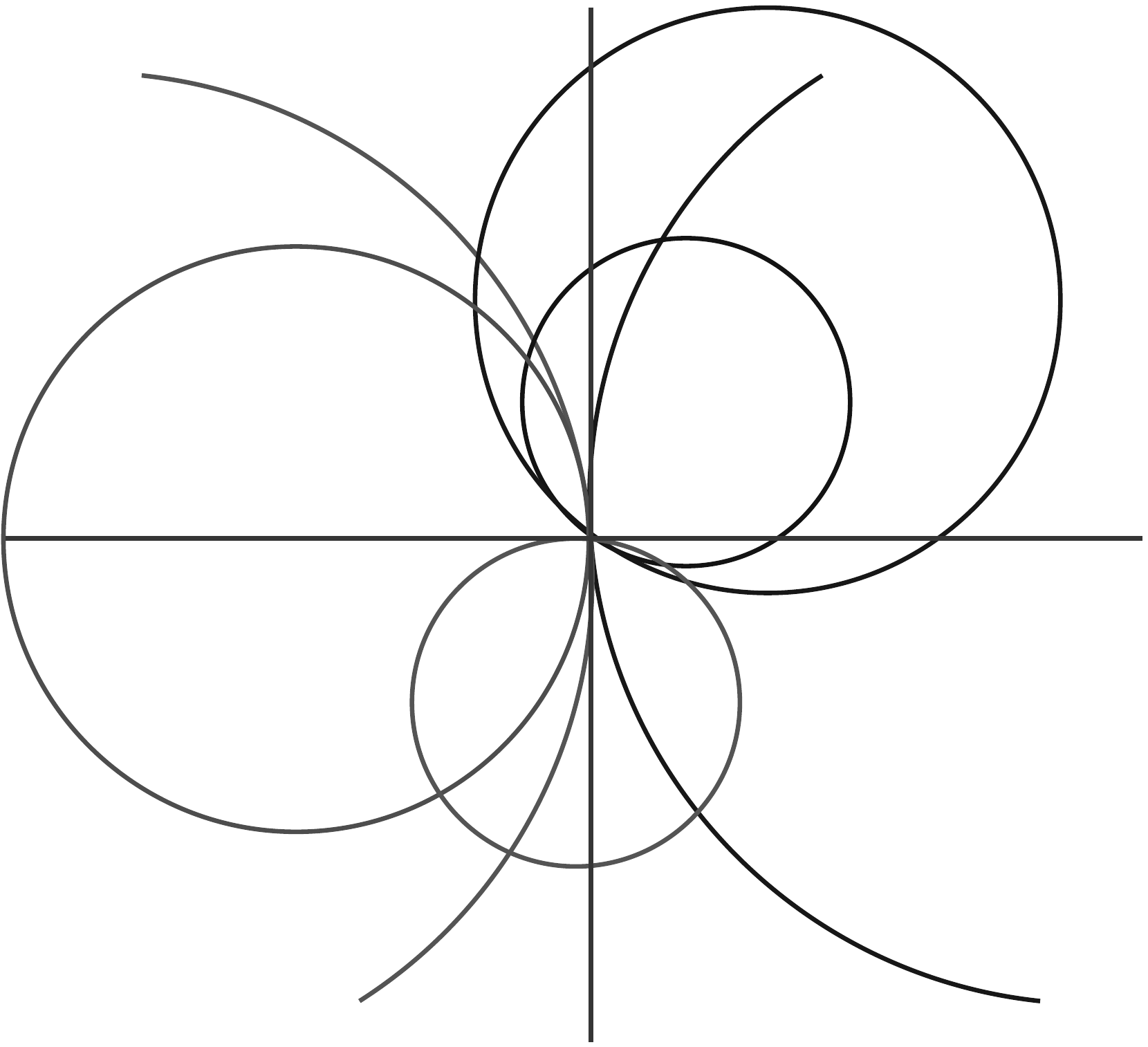}
\end{center}
\caption{Tomography by circles. }
\label{fig:circles}
\end{figure}

The conformal inversion 
\begin{equation}\label{eq:conformal}
(x,y)=\varphi(q,p)=\left(\frac{q}{q^2+p^2},\frac{p}{q^2+p^2}\right),
\end{equation}
maps the family of  lines
\begin{equation}
\lambda  -\mu x - \nu y = 0
\end{equation}
into a family of circles
\begin{equation}
\lambda  (q^2+p^2) -\frac{\mu}{q}  - \nu p = 0,
\label{eq:circles}
\end{equation}
centered at
\begin{equation}\label{eq:centers}
\left(\frac{\mu}{2\lambda },\frac{\nu}{2\lambda } \right)
\end{equation}
and passing through the origin (see Fig. \eref{fig:circles}). When $\lambda =0$ they degenerate into
lines through the origin.

The Jacobian of the transformation
\begin{equation}
J(q,p)=\left|\frac{\partial(x,y)}{\partial(q,p)}\right|=\frac{1}{(q^2+p^2)^2},
\end{equation}
does never vanish as a consequence of the fact that {\boldmath$\varphi$} is a 
 autodiffeomorphism of $\R^2-\{(0,0)\}$. The origin $(0,0)$ 
is irrelevant for  tomographic integral transformations because
the singularity only affects to a zero measure set. 

Eqs.~(\ref{eq:radondefns})-(\ref{eq:invgens}) then become
\begin{eqnarray}
& & \!\!\!\!\!\!\!\!\!\!\!\!\!\!\!\!\!\!\!\!\!\!\!\!\widehat{f}(\lambda ,\mu,\nu) = \left\langle
\delta\left(\lambda -\frac{\mu q}{q^2+p^2} -\frac{\nu p}{q^2+p^2}\right)\right\rangle \nonumber\\
& & \!\!\!\!\!\!\!\!\!\!\!\!\!\!\!\!\!\!\!\!\!\!\!\!= \int_{\mathbb{R}^2}dq\, dp\,  f(q,p) \delta\left(\lambda -\frac{\mu q}{q^2+p^2} -\frac{\nu p}{q^2+p^2}\right) 
\label{eq:radondefnc}
\end{eqnarray}
and
\begin{eqnarray}
f(q,p) = \int_{\mathbb{R}^3} {d\lambda \,
d\mu\, d\nu} \, \widehat{f}(\lambda ,\mu)
\frac{e^{i(\lambda -\frac{\mu q}{q^2+p^2} -\frac{\nu
p}{q^2+p^2})}}{{(2\pi)^2}(q^2+p^2)^2}\,
. \quad
\label{eq:invgenc}
\end{eqnarray}

\subsection{Tomography by hyperbolas in the plane}
\label{sec-hyperbolas}

\begin{figure}[t]
\begin{center}
\includegraphics[width=8cm]{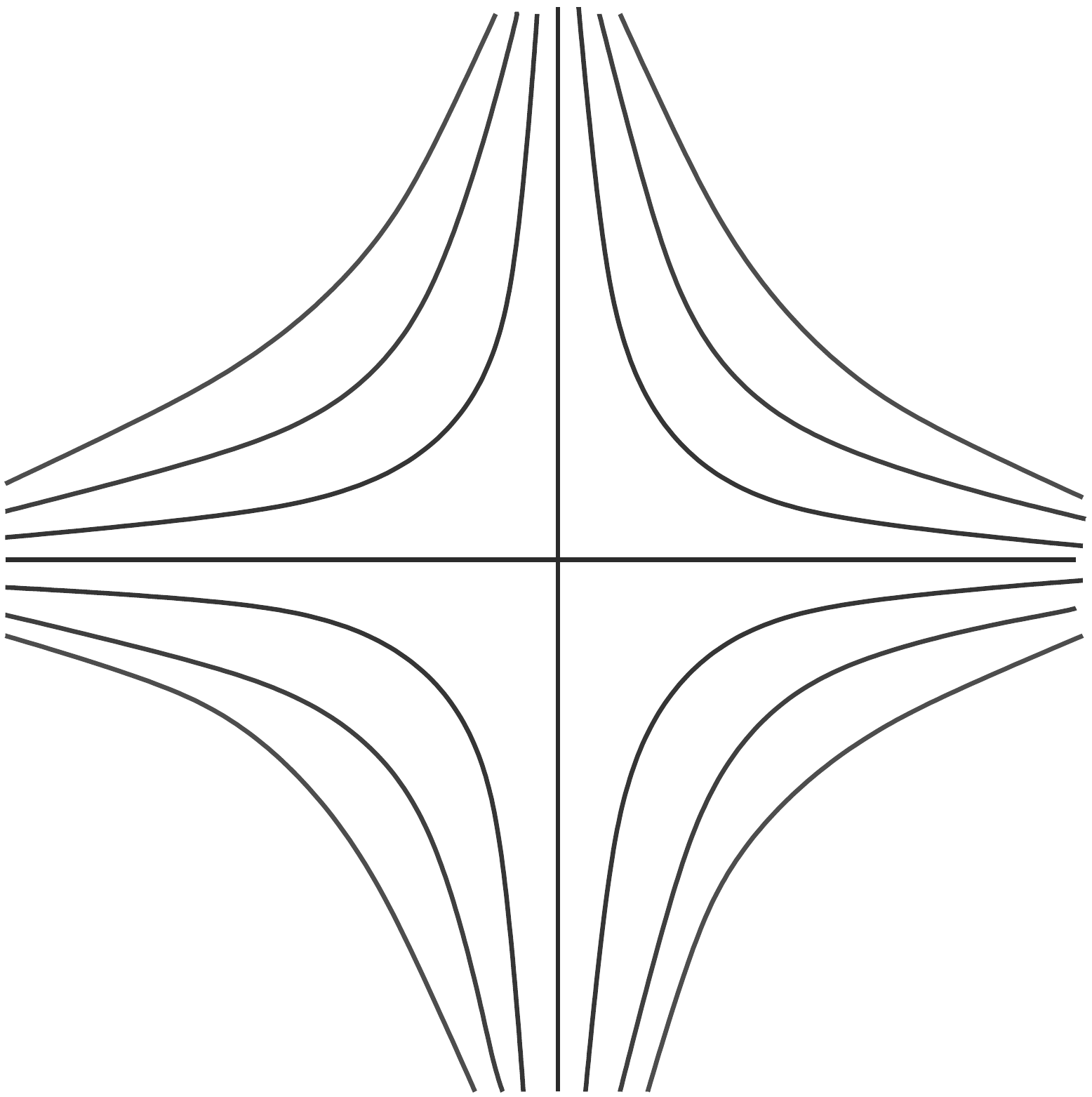}
\end{center}
\caption{Tomography by hyperbolas induced by the map $\varphi$ \eref{eq:inversion}. }
\label{fig:hyperb}
\end{figure}

The   family of  lines 
\begin{equation}
\lambda  -\mu x - \nu y = 0
\end{equation}
on the plane
is mapped into a family of hyperbolas
\begin{equation}
\lambda  -\frac{\mu}{q}  - \nu p = 0,
\label{eq:hyperbola}
\end{equation}
with asymptotes
\begin{equation}
q= 0, \qquad p= \frac{\lambda }{\nu}
\end{equation}
by the transformation
\begin{equation}\label{eq:inversion}
(x,y)=\varphi(q,p)=\left(\frac{1}{q},p\right).
\end{equation}
For $\mu>0$ the hyperbolas are in the second and fourth quadrants,
while for $\mu<0$ they are in the first and third quadrants (see Fig. \eref{fig:hyperb}). When
$\mu=0$ or $\nu=0$ they degenerate into lines respectively
horizontal or vertical.

The Jacobian 
\begin{equation}
J(q,p)=\left|\frac{\partial(x,y)}{\partial(q,p)}\right|=\frac{1}{q^2},
\end{equation}
does not vanish again because $\varphi$ is  diffeomorphism in  $\R^2-\{(0,y), y\in \R\}$.

Eqs.~(\ref{eq:radondefns})-(\ref{eq:invgens}) become
\begin{eqnarray}
\!\!\widehat{f}(\lambda ,\mu,\nu) &=& \left\langle \delta\left(\lambda -\frac{\mu}{q} -\nu p\right)\right\rangle \nonumber\\
& =& \int_{\mathbb{R}^2}dq\, dp \,  f(q,p) \delta\left(\lambda -\frac{\mu}{q} -\nu
p\right) 
\label{eq:radondefnh}
\end{eqnarray}
and
\begin{eqnarray}
f(q,p) = \int_{\mathbb{R}^3} \frac{d\lambda d\mu\, d\nu}{(2\pi)^2} \, \widehat{f}(\lambda ,\mu) \frac{1}{q^2}
e^{i(\lambda -\frac{\mu}{q}-\nu p)} . \quad
\label{eq:invgenh}
\end{eqnarray}

\subsection{Hyperbolic tomography  in $\R^n$}

It is possible to generalize this kind of hyperbolic tomography in  terms of quadratic forms. 
Let us consider the following tomographic map
\begin{eqnarray}
\!\!\! \widehat{f}(\boldsymbol{\xi},\nu,\lambda )\! =\!\! \int {d^n  \mathbf{p}\ d^n  \mathbf{q} }\,   \delta \left(\lambda -{\boldsymbol{\xi}}\cdot  \mathbf{q}-\nu (\mathbf{p}, \mathbf{q}) \right)
 f( \mathbf{p}, \mathbf{q}),
\nonumber
\end{eqnarray}
where $ \mathbf{p}$ and $ \mathbf{q}$ are vectors in $\R^n$ and
\begin{equation}
\nu ( \mathbf{p}, \mathbf{q})=\sum_{j=1}^n q_j \nu_j p_j.
\end{equation}
is a bilinear non-degenerate diagonal form defined by a vector $\boldsymbol{\nu}=(\nu_1,\nu_2,\cdots,\nu_n)\in \R^n$.

This map corresponds to the deformation of standard multidimensional Radon transform
by means of the following local diffeomorphism
\beq
\boldsymbol{\varphi}(q_i,p_j)= \left({q_i}, {q_j} p_j\right),
\label{diffeo2}
\eeq
whose Jacobian is
\beq
J( \mathbf{q}, \mathbf{p})=\left|\frac{\partial(\mathbf{x},\mathbf{y})}{\partial( \mathbf{q}, \mathbf{p})}\right|=\prod_{j=1}^n| q_j |
\eeq

Thus,  the inverse map is given by
\begin{equation}\label{eq:inv2}
 f( \mathbf{p}, \mathbf{q})  = \frac1{(2\pi)^{2n}}\int d^n \boldsymbol{\xi}\  d^n\boldsymbol{\nu}\  d\lambda \, 
\widehat{f}(\boldsymbol{\xi}, \boldsymbol{\nu},\lambda ) \,\prod_{j=1}^n| q_j |\,  {\rm e}^{i\left(\lambda -{\boldsymbol{\xi}}\cdot
 \mathbf{q}-\nu (\mathbf{p}, \mathbf{q}) \right)} .
\end{equation}
as can be derived from integration over $\lambda $ 
\begin{eqnarray}
 f( \mathbf{p}, \mathbf{q}) &=& \frac1{(2\pi)^{2n}}\int {d^n \boldsymbol{\xi}\  d^n \boldsymbol{\nu}\  {d^n  \mathbf{p}'\ d^n  \mathbf{q}' }\,} f( \mathbf{p}', \mathbf{q}')
\nonumber\\
& & \times \prod_{j=1}^n| q_j |\, {\rm e}^{i\left(\boldsymbol{\xi}\cdot q'+\nu
 (\mathbf{p}' , \mathbf{q}')-\boldsymbol{\xi}\cdot  \mathbf{q}-\nu
 (\mathbf{p}, \mathbf{q}) \right)}  \nonumber\\
&=&\int {  d^n \boldsymbol{\nu}\   {d^n  \mathbf{p}' }\,} \prod_{j=1}^n\left|
\frac{q_j}{2\pi}\right |\,  {\rm e}^{i\nu ( \mathbf{p}'- \mathbf{p}, \mathbf{q})}f( \mathbf{p}', \mathbf{q}) \nonumber\\
&=& \int { {d^n  \mathbf{p}' }\,} \delta( \mathbf{p}'- \mathbf{p}) f( \mathbf{p}', \mathbf{q}).
\label{eq:inv3}
\end{eqnarray}
This corresponds to the higher dimensional generalization of the
Bertrand-Bertrand tomography \cite{Ber-Ber}

\begin{figure}[t]
\begin{center}
\includegraphics[width=8cm]{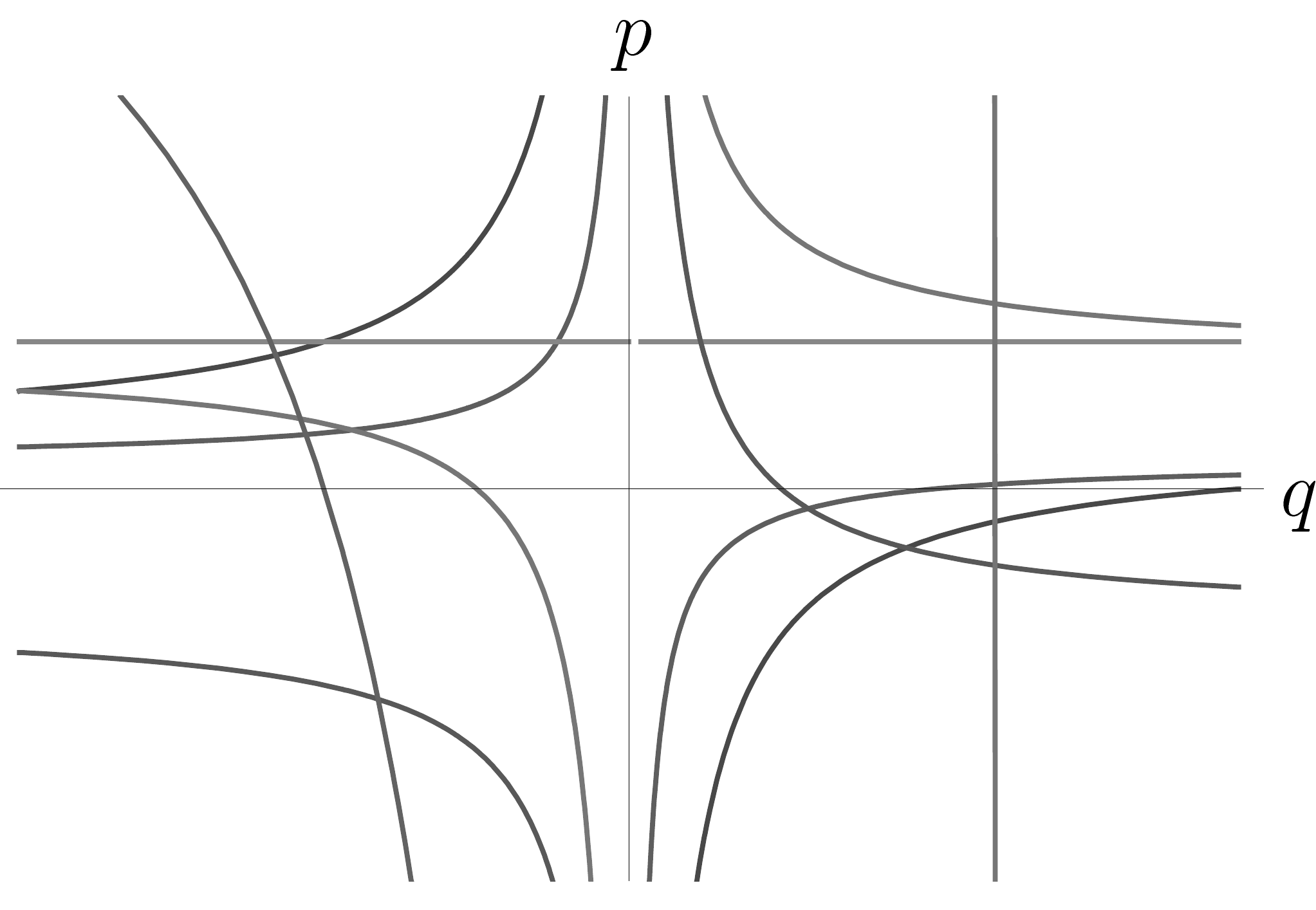}
\end{center}
\caption{Hyperbolic tomography induced by the map $\boldsymbol{\varphi}$ defined by Eq. \eref{diffeo2} . }
\label{fig:hyperc}
\end{figure}

{Notice that the distribution of hyperboloids in the plane when $n=2$
is quite different from
those analyzed in the previous subsection (see Fig. \eref{fig:hyperc}).}

\subsection{Tomography by quadrics}

We remark that  the above generalizations  of X rays tomograms  involve integration over
 unbounded  submanifolds. It will be interesting to generalize
the Radon transform  over compact submanifolds.
 
This can be achieved by shifting a given pattern of quadric like
\begin{equation}
\label{eq:quartic}
\lambda =(\mathbf{q}-\mu) \cdot B ( \mathbf{q}-\mu))+\mathbf{a}\cdot( \mathbf{q}-\mu)
\end{equation}
where $B$ is a non-degenerate symmetric operator and $\mathbf{a}, \boldsymbol{\mu}\in\R^n$. A new 
 tomographic map can be defined by
\begin{eqnarray}
 \widehat{f}_{\mathbf{a} B }(\lambda ,&&\!\!\!\!\!\!\!\!\!\!\!\boldsymbol{\mu})  = \int {d^n  \mathbf{q} \ }\, f( \mathbf{q})\nonumber\\
&\times& \delta \left(\lambda -( \mathbf{q}-\boldsymbol{\mu})\cdot B ( \mathbf{q}-\boldsymbol{\mu})-\mathbf{a}\cdot(\mathbf{q}-\boldsymbol{\mu})\right). 
\label{new}
\end{eqnarray}
When $B=\boldsymbol{\mu}=0$ the transform (\ref{new}) reduces to the standard Radon 
transform. But in general (\ref{new}) defines a completely different type of transform
supported on the quadrics defined by equation (\ref{eq:quartic}).
 It is easy to show that the inverse map is given by
\begin{eqnarray}
 f( \mathbf{q})  &=&\displaystyle\frac{|\det B|}{\pi^n } \int {d^n \boldsymbol{\mu}\  d\lambda }  \,  \widehat{f}_{\mathbf{a} B }(\lambda ,\boldsymbol{\mu})\nonumber\\
 &\times& \displaystyle{\rm e}^{i\left(\lambda -( \mathbf{q}-\boldsymbol{\mu})\cdot B ( \mathbf{q}-\boldsymbol{\mu}))-\mathbf{a}\cdot (\mathbf{q}-\boldsymbol{\mu})\right)}.
\label{eq:inv}
\end{eqnarray}
This can shown by plugging  the definition of the tomographic map \eref{new} into \eref{eq:inv}
(\ref{eq:inv})
\begin{eqnarray}
& &\!\!\!\!\!\!\!\!\!\! \displaystyle\frac{|\det B|}{\pi^n }\int {d^n \boldsymbol{\mu}\  d\lambda }\,  {\rm
e}^{i\left(\lambda -( \mathbf{q}-\boldsymbol{\mu})\cdot B ( \mathbf{q}-\boldsymbol{\mu}))-\mathbf{a}\cdot (\mathbf{q}-\boldsymbol{\mu})\right)}  \widehat{f}_{\mathbf{a} B }(\lambda ,\boldsymbol{\mu}) \nonumber\\ 
&&= \displaystyle\frac{|\det B|}{\pi^n }\int {d^n \boldsymbol{\mu}\  d\lambda }\,  {\rm
e}^{i\left(\lambda -( \mathbf{q}-\boldsymbol{\mu})\cdot B ( \mathbf{q}-\boldsymbol{\mu}))-\mathbf{a}\cdot (\mathbf{q}-\boldsymbol{\mu})\right)}
 \nonumber\\
& & \times \int {d^n \boldsymbol{\xi}\ } \delta\left(\lambda -(\boldsymbol{\xi}-\boldsymbol{\mu})\cdot B
(\boldsymbol{\xi}-\boldsymbol{\mu})-\mathbf{a}\cdot(\boldsymbol{\xi}-\boldsymbol{\mu})\right) f(\boldsymbol{\xi})\nonumber,
\end{eqnarray}
which integrating over $\lambda $  gives
\begin{eqnarray}
 &&\displaystyle\frac{|\det B|}{\pi^n }\int {d^n \boldsymbol{\xi}\ }f(\boldsymbol{\xi})  \nonumber\\
& &  \times \int {d^n \boldsymbol{\mu}}\, {\rm
e}^{i\left((\boldsymbol{\xi}-\boldsymbol{\mu})\cdot B (\boldsymbol{\xi}-\boldsymbol{\mu})-( \mathbf{q}-\boldsymbol{\mu})\cdot B ( \mathbf{q}-\boldsymbol{\mu})-\mathbf{a}\cdot (\mathbf{q}-\boldsymbol{\xi})\right)}
\nonumber\\
&=&\frac{|\det B|}{\pi^n }\int {d^n \boldsymbol{\xi}\ }{d^n \boldsymbol{\mu}}\,f(\boldsymbol{\xi}) {\rm e}^{i\left[\boldsymbol{\xi}\cdot B \boldsymbol{\xi}- \mathbf{q}\cdot B  \mathbf{q}+\left( \mathbf{q}-\boldsymbol{\xi}\right)\cdot 2 B (\boldsymbol{\mu}-{B^{-1}a/2})\right]} 
\nonumber\\
&=&\int {d^n \boldsymbol{\xi}\ }f(\boldsymbol{\xi}) {\rm e}^{i\left[\boldsymbol{\xi}\cdot B \boldsymbol{\xi}- \mathbf{q}\cdot B
 \mathbf{q})\right]}\delta^n\left( \mathbf{q}-\boldsymbol{\xi}\right)=f( \mathbf{q}). \nonumber
\end{eqnarray}

\begin{figure}[t]
\begin{center}
\includegraphics[width=4cm]{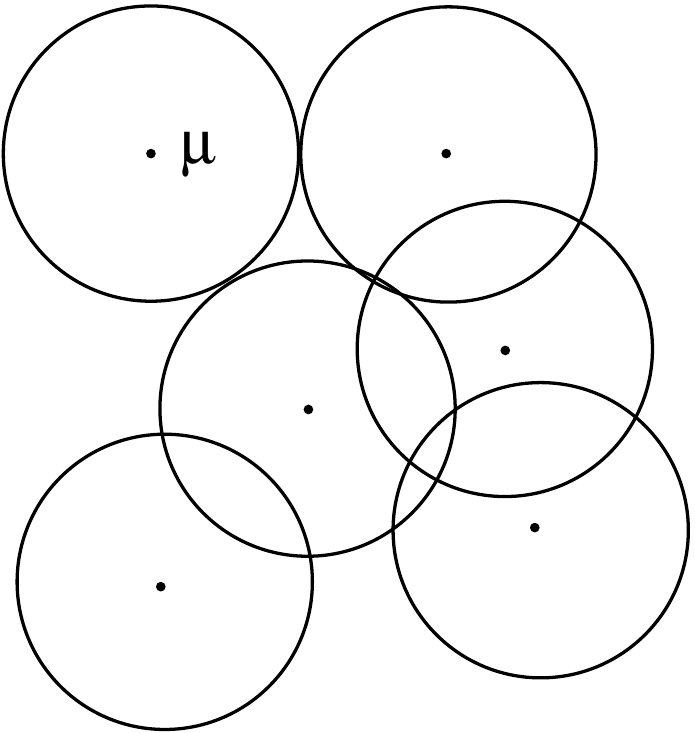}
\end{center}
\caption{Tomography on circles in  the plane for the Hall effect.
}
\label{fig:cylinder}
\end{figure}

The meaning of the above tomographic map depends on the physical 
character of B. If we assume that B is strictly positive (elliptic case),
the map corresponds to averages of $f$ along the ellipsoids
defined by the equation (\ref{eq:quartic}). In particular if $a=0$ and all
eigenvalues of $B$ are equal to $b^2$ it corresponds to
integration over spheres centered at $\boldsymbol{\mu}$ (see Fig. \eref{fig:cylinder}) 
\begin{equation}\label{eq:spheres}
b^2 (\mathbf{q}-\boldsymbol{\mu})^2=\lambda.
\end{equation}

In the two dimensional case corresponds to trajectories of
particles moving under the action of a constant magnetic field on
a plane (Hall effect).

 In the case that $B$ has positive and negative eigenvalues
this correspond to the hyperbolic tomography which averages of $f$
along the hyperboloids defined by the equation (\ref{eq:quartic}),
e.g.,
\begin{equation}\label{eq:hyperboloids}
b^2(q_1-\mu_1)^2-c^2(q_2-\mu_2)^2=\lambda .
\end{equation}

In the case of degenerated $B$ forms we have an hybrid transform. $B$ can
 be decomposed into a
non-degenerated bilinear form and  a linear form. In that case the
tomography of the linear components should be treated as the standard
Radon transform, whereas the non-degenerate
variables should  transform as above.

{Notice that the distribution of circles is quite different from
those obtained by the Radon transform defined by rescaling in subsection \eref{sec-circles}.
In that case the centers and the
radius changed from one circle to another, whereas here all the 
circles are of fixed radius and only the center of the
circles are varying by shifts. The new tomographic map makes possible a local
implementation of tomography because the marginals
involved into the resconstruction of $f(p,q)$ only involve
integrations in compact domain around the point $(p,q)$.}
The trajectories involved  in the  construction  the tomogram are
all trajectories of a charged particle with the same energy moving in a constant 
magnetic field. 
This points out the possible applications to tomography by charged
particles moving on a transversal magnetic fields over samples with
smaller sizes than the cyclotron radius $mv/eB$.

Finally we remark that the symplicity of the inverse formula seems to be related to the integrable
character of the underlying dynamical system whose  ray trajectories are  involved in the construction of
the tomogram.
We conjecture that this a generic issue, i.e. tomograms defined by phase space
trajectories of integrable systems have simple analytic inversion formulas. 


\section{Tomography and coherent states}\label{sec:coherent}
There are several representations of quantum states providing the
possibility to present equivalent, but different in their form, formulations
of quantum mechanics \cite{AmJPhys}. Among them, the Wigner quasi-distribution function \cite{Wi32} and the well known Husimi-Kano $K-$function \cite{Husimi1940,Kano56}, (in this paper we have decided to keep up with the original notations of the pioneer papers on the subject).
Quasi-distributions are usually referred to as phase-space representations
of quantum states. Another important phase-space representation is related
to the Sudarshan's $\phi$ diagonal coherent-state representation \cite{Sudarshan1963,Glauber1963}. 
Interestingly, the last two phase space representations,  $K$ and $\phi$, are based on the use of coherent states 
and allow for an interpretation in a tomographic scheme which fits nicely in the general description given above.

In the quantum case, tomography may also be casted in a quantizer-dequantizer scheme. In this sense, the Husimi-Kano
function is a dequantization, and the Sudarshan's $\phi$ is just its dual symbol. So, the  coherent-state tomographic point of view provides a unifying description that encompasses also the so called photon number tomography. Here, we limit to 
the coherent-state (CS) tomography for a single degree of freedom, and we adopt a pedagogical style which is reminiscent of many joint papers with Volodya Man'ko.

\subsection{Coherent states: main properties}
In the case of a single degree of freedom, the Weyl map associates with any point $(x,\alpha)$ of the phase-space of the system a unitary operator $\hat{W}(x,\alpha)$ acting on a Hilbert space $\mathcal{H},$ which can be realized in terms of square integrable functions $\psi(x)$ \cite{newmarmobook}. Upon switching to complex numbers $z=(x+i \alpha)/\sqrt{2}$, we recognize that $\hat{W}(x,\alpha)$ is nothing but the usual displacement operator 
\begin{equation}
\hat{\mathcal{D}}\left( z \right) =\exp \left( z \hat{a}^{\dagger }-z ^{\ast }\hat{%
a}\right) ,
\end{equation}
which acting on the vacuum Fock state $\vert 0 \rangle$ , $\hat{a}\vert 0 \rangle=0,$ generates the coherent state $\vert z \rangle,$ where
\begin{equation}
\left\vert z \right\rangle =\exp (-\frac{\left\vert z \right\vert ^{2}}{2}%
)\exp \left( z \hat{a}^{\dagger }\right) \exp \left( -z ^{\ast }\hat{a}%
\right) \left\vert 0\right\rangle =\exp (-\frac{\left\vert z \right\vert ^{2}%
}{2})\sum_{j=0}^{\infty }\frac{z ^{j}}{j!}\hat{a}^{\dagger j}\left\vert
0\right\rangle \, .
\end{equation}

We recall that the coherent states are a (over-) complete set in the Hilbert
space $\mathcal{H}.$ Any bounded set containing a limit point $z_{0}$ in the
complex $z-$plane defines a complete set of coherent states containing a
limit point, the coherent state $\left\vert z_{0}\right\rangle ,$ in the
Hilbert space $\mathcal{H}$. In particular, any Cauchy sequence $\{z_{k}\}$
of complex numbers defines a Cauchy sequence of coherent states $\left\{
\left\vert z_{k}\right\rangle \right\} ,$ which is a complete set. The same
holds for any extracted subsequence. This completeness property holds as $%
\exp \left( \left\vert z\right\vert ^{2}/2\right) \left\langle z|\psi
\right\rangle $ is an entire analytic function of the complex variable $%
z^{\ast },$ for any $\left\vert \psi \right\rangle \in \mathcal{H}$. Then
\begin{equation}
\left\langle z_{k}|\psi \right\rangle =0\quad \forall k\Rightarrow
\left\vert \psi \right\rangle =0 \, ,
\end{equation}
because $z_{0}^{\ast }$ is a non-isolated zero of an analytic function.

Besides, any bounded operator $\hat{A}$ may be completely reconstructed from its
diagonal matrix elements $\left\langle z _{k}\left\vert \hat{A}\right\vert z
_{k}\right\rangle .$ In fact, $\exp \left( \left\vert z \right\vert
^{2}/2+\left\vert z^{\prime}\right\vert ^{2}/2\right) \left\langle z
\left\vert \hat{A} \right\vert z^{\prime}\right\rangle $ is an analytical function
of the complex variables $z ^{\ast },z^{\prime},$\ so it is uniquely
determined by its value $\exp \left( \left\vert z \right\vert ^{2}\right)
\left\langle z \left\vert \hat{A}\right\vert z \right\rangle $ on the diagonal $%
z^{\prime}=z .$ This is an entire function of the real variables $\Re z ,\Im
z ,$ which is in turn uniquely determined by its values on any set with an
accumulation point.

\subsection{Tomographic sets in infinite dimensional spaces}
The nice properties of the coherent states previously discussed are the ground on which the CS tomography can be builded.
To this aim, we need to introduce a linear space $V$ and its dual $V^{\ast}.$ 

Preliminarly, consider an abstract case. Suppose we have chosen in $V^{\ast}$ the tomographic set of observables $\mathcal{N}$ parametrzed as $\left\{
\hat{P}_{\nu }\right\} , {\nu \in I}$, where $\nu$ is a multi-index belonging to a set of indices $I$. In the infinite dimensional case there are several relevant spaces that one could choose as $V$, like the space of
bounded operators $B(\mathcal{H})$ and that of compact operators $C(\mathcal{%
H})$, the space of Hilbert-Schmidt operators $\mathfrak{I}_{2}$ and that of
trace-class operators $\mathfrak{I}_{1}$. Their mutual relations are:
\begin{equation}
\mathfrak{I}_{1}\subset \mathfrak{I}_{2}\subset C(\mathcal{H})\subset B(%
\mathcal{H}).
\end{equation}
$B(\mathcal{H})$ $($and $C\mathbb{(\mathcal{H}))}$ are Banach spaces, with
the norm $\left\| \hat{A}\right\| =\sup_{(\left\| \psi \right\| =1)}\left\| \hat{A}\psi
\right\| ,$ while $\mathfrak{I}_{2}$ is a Hilbert space with scalar product $%
\left\langle \hat{A}|\hat{B}\right\rangle =\mathrm{Tr}\left( \hat{A}^{\dagger }\hat{B}\right) $.
Finally, $\mathfrak{I}_{1}$ is a Banach space with the norm $\left\|
\hat{A}\right\| _{1}=\mathrm{Tr}\left( \left| \hat{A}\right| \right) .$ The following
inequalities hold true:
\begin{equation}
\left\| \hat{A}\right\| \leq \left\| \hat{A}\right\| _{2}\leq \left\| \hat{A}\right\| _{1}.
\end{equation}
So $\mathfrak{I}_{2}$, the only Hilbert space at our disposal to implement
our definition of tomographic set, is endowed with a topology which, when
restricted to the trace-class operators, is not equivalent to the topology
of $\mathfrak{I}_{1}$. Similar problems show up in the identification of
states and maps in infinite dimensions \cite{MaKuGra}.

However, as any density state is associated with a trace class operator $\hat{\rho},$ the natural choice is to identify the space $V$ with 
$\mathfrak{I}_{1}$. Then, recalling that $%
\mathfrak{I}_{1}$ is a $\ast -$ideal in its dual space $B(\mathcal{H})$:
\begin{equation}
\mathfrak{I}_{1}^{\ast }=B(\mathcal{H}),
\end{equation}
 we have $V^{\ast}=B(\mathcal{H})$. So, the tomographic map $F_{\hat{\rho}}$ defined on the tomographic set $\mathcal{N}=\left\{%
\hat{P}_{\nu }\right\} , {\nu \in I},$ is just the value of the linear functional $\mathrm{Tr}\left( \hat{P}_{\nu } \cdot \right) $ in $\hat{\rho.}$
In other words, we can write the tomographic pairing between states and observables as 
 \begin{equation}
\mu(\hat{P}_{\nu },\hat{\rho}):=F_{\hat{\rho}}\left( \hat{P}_{\nu }\right):=\mathrm{Tr}\left( \hat{P}_{\nu } \hat{\rho} \right) 
\end{equation}
The invertibility of the map $F_{\hat{\rho}}$, or, equivalently, the full reconstruction of any density state $\hat{\rho}$ from its tomograms is guaranteed  if we choose a tomographic set of trace class observables, e.g., rank-one projectors, complete in $\mathfrak{I}_{1}.$ The completeness of the tomographic set yields
\begin{equation}
F_{\hat{\rho}}\left( \hat{P}_{\nu }\right)=\mathrm{Tr}\left( \hat{P}_{\nu } \hat{\rho} \right) =0 \quad \forall \nu \in I  \Longrightarrow \hat{\rho}=0. 
\end{equation}

The choice of tomographic sets as complete sets of rank-one projectors gives the possibility to extend the tomographic representation to observables or, in general, to bounded operators. To do that, we have to interchange the role of states and observable. We can use the self-duality of the Hilbert-Schmidt space $\mathfrak{I}_{2}.$ So, the tomographic set is a  set of density states, complete in $\mathfrak{I}_{2}$, and the invertibility of the tomographic map associated to a bounded operator $\hat{A}$ reads
\begin{equation}
F_{\hat{A}}\left( \hat{P}_{\nu }\right):=\left\langle \hat{P}_{\nu }|\hat{A}\right\rangle = \mathrm{Tr}\left( \hat{P}_{\nu } \hat{A} \right)=0 \quad \forall \nu \in I  \Longrightarrow \hat{A}=0\quad \&\quad \hat{A}\in \mathfrak{I}_{2}  \label{Az}
\end{equation}
Then, as $\mathfrak{I}_{2}$ is a $\ast -$ideal in $B(\mathcal{H}),$ there
may exist a non-zero operator $\hat{B},$ which is bounded but not
Hilbert-Schmidt, such that
\begin{equation}
\mathrm{Tr}\left( \hat{P}_{\nu }B\right) =0\quad \forall \nu \in I
\end{equation}
So, different operators may be tomographically separated only
when their difference is Hilbert-Schmidt. In other words, by choosing $V=V^{\ast}=\mathfrak{I}_{2}$ one cannot  get tomographic representations of bounded, not Hilbert-Schmidt operators.. Nevertheless, there exists the possibility of such a representationn, when the set $\left\{ \hat{P}_{\nu }\right\} $ of rank-one projectors is complete even in $\mathfrak{I}_{1}$. Then, recalling that $%
\mathfrak{I}_{1}$ is a $\ast -$ideal in its dual space $B(\mathcal{H})$, the expression $\mathrm{Tr}\left( P_{\mu }A\right) $ is nothing but the
value of the linear functional $\mathrm{Tr}\left( \cdot \hat{ A}\right) $ in $%
\hat{P}_{\nu }$ and the invertibility condition hods without constraints:
\begin{equation}
\mathrm{Tr}\left( \hat{P}_{\nu }A\right) =0\quad \forall \nu \in I \Longrightarrow
0=\left\| \mathrm{Tr}\left( \cdot A\right) \right\| =\left\| A\right\|
\Longrightarrow A=0.
\end{equation}

So, any bounded operator can have a tomographic representation based on sets of rank-one projectors
which are complete both in $\mathfrak{I}_{2}$ and in $\mathfrak{I}_{1}$. As
it turns out, this is the case for the main tomographic sets. The rank-one projectors $\hat{P}_{z}=\vert z \rangle \langle z \vert$ associated with a complete set of coherent states are
complete in the Hilbert space $\mathfrak{I}_{2}$. In particular, any Cauchy
sequence $\left\{ \left\vert z _{k}\right\rangle \right\} $ generates a
tomographic set $\left\{ \left\vert z _{k}\right\rangle \left\langle z
_{k}\right\vert \right\} $. In fact, bearing in mind the previous remark on
the reconstruction of a bounded operator, it results
\begin{equation}
\mathrm{Tr}(A\left\vert z _{k}\right\rangle \left\langle z _{k}\right\vert
)=\left\langle z _{k}\left\vert A\right\vert z _{k}\right\rangle =0\quad
\forall k\Rightarrow A=0\quad \&\quad A\in B(\mathcal{H}).
\end{equation}
This shows that a tomographic set of coherent state projectors is complete
even in $\mathfrak{I}_{1}.$  

\subsection{The coherent state tomography}
Hereafter, we address the case when $z$ varies
in the whole complex plane, and call CS tomography  only the tomography based on the whole set of rank-one projectors $\{P_{z}\}=\left\{\vert z \rangle \langle z \vert \right\},{z \in \mathbb{C}}$.

It is possible to interpret the well known Husimi-Kano $K$-symbol of a
(bounded) operator $\hat{A}$ as the CS tomographic representation of $\hat{A}$%
:
\begin{equation}
K_{A}(z):=\left\langle z\left| \hat{A}\right| z\right\rangle =:\mathrm{Tr}%
(\left| z\right\rangle \left\langle z\right| \hat{A}).  \label{HK1}
\end{equation}
In particular, when $\hat{A}$ is chosen as a density operator $\hat{\rho},$
the identity holds
\begin{equation}
\int \frac{d^{2}z}{\pi }\left\langle z\left| \hat{\rho}\right|
z\right\rangle =\mathrm{Tr}(\hat{\rho})=1,
\end{equation}
which allows for the probabilistic interpretation of the CS tomography. As a
matter of fact \cite{SudKlaubook} the $K$-symbol exists also for a number of
non-bounded operators. The CS tomographic set is complete both in $\mathfrak{%
I}_{2}$, the space of Hilbert-Schmidt operators, and in $\mathfrak{I}_{1},$
the space of trace class operators acting on the space of states. In fact,
the formulae
\begin{equation}
\hat{A}=\int \frac{d^{2}z}{\pi }\frac{d^{2}z^{\prime }}{\pi }\left\langle
z\left| \hat{A}\right| z^{\prime }\right\rangle \left| z\right\rangle
\left\langle z^{\prime }\right|
\end{equation}
and \cite{MandelWolf}
\begin{equation}
\left\langle z\left| \hat{A}\right| z^{\prime }\right\rangle =\mathrm{e}^{-%
\frac{\left| z\right| ^{2}+\left| z^{\prime }\right| ^{2}}{2}%
}\sum_{n,m=0}^{\infty }\frac{(z^{\ast })^{n}(z^{\prime })^{m}}{n!m!}\left[
\frac{\partial ^{n+m}}{\partial z^{\ast n}\partial z^{m}}\left( \mathrm{e}%
^{\left| z\right| ^{2}}\left\langle z\left| \hat{A}\right| z\right\rangle
\right) \right] _{_{z=0}^{z^{\ast }=0}}  \label{19}
\end{equation}
show that if the tomograms $\left\langle z\left| \hat{A}\right|
z\right\rangle $ of a bounded operator $\hat{A}$ vanish for any $z\in
\mathbb{C}$ , then $\hat{A}$ is the zero operator. 

So, a resolution of the unity exists, which allows for the full
reconstruction of any density state or (bounded) operator from its CS tomograms. We are
interested in the explicit determination of such a formula. Now, the
Sudarshan's diagonal coherent state representation $\phi_{A}(z)$ of an
operator $\hat{A}$ is defined through the equation
\begin{equation}
\hat{A}=\int \frac{d^{2}z}{\pi }\phi_{A}(z)\left| z\right\rangle
\left\langle z\right| .  \label{Defk}
\end{equation}
So, to get the tomographic reconstruction formula we have to invert the
well-known relation
\begin{equation}
K_{A}(z^{\prime })=\left\langle z^{\prime }\left| \hat{A}\right| z^{\prime
}\right\rangle =\int \frac{d^{2}z}{\pi }\phi_{A}(z)\left| \left\langle
z|z^{\prime }\right\rangle \right| ^{2}=\int \frac{d^{2}z}{\pi }\phi_{A}(z)%
\mathrm{e}^{-\left| z-z^{\prime }\right| ^{2}}  \label{Fourtras}
\end{equation}
which follows at once from Eq.(\ref{Defk}) defining $\phi_{A}(z)$. This
relation shows that $K_{A}(z^{\prime })$ is given by the convolution product
of $\phi_{A}$ times a gaussian function. Then, denoting with $%
K_{A}(u^{\prime },v^{\prime })$ and $\phi_{A}(u,v)$ the $K$
and $\phi$ symbols, with $z^{\prime }=u^{\prime }+iv^{\prime }$ and $%
z=u+iv$, the Fourier transform \cite{Schleich} of Eq.(\ref{Fourtras}%
) reads:
\begin{eqnarray}
&&\int \frac{du^{\prime }dv^{\prime }}{2\pi }K_{A}(u^{\prime
},v^{\prime })\mathrm{e}^{-i(\xi u^{\prime }+\eta v^{\prime })}=
\\
&&\int \frac{du^{\prime }v^{\prime }}{2\pi }\int \frac{dudv%
}{\pi }\phi_{A}(u,v) \mathrm{e}^{-\left( u-u^{\prime }\right)
^{2}-\left( v-v^{\prime }\right) ^{2}}\mathrm{e}^{-i(\xi u^{\prime }+\eta v^{\prime })}  \nonumber
\end{eqnarray}
and we readily obtain
\begin{equation}
\tilde{K}_{A}(\xi ,\eta )=\mathrm{e}^{-(\xi ^{2}+\eta ^{2})/4}\tilde{\phi}%
_{A}(\xi ,\eta ),
\end{equation}
from which
\begin{equation}
\tilde{\phi}_{A}(\xi ,\eta )=\mathrm{e}^{(\xi ^{2}+\eta ^{2})/4}\tilde{K}%
_{A}(\xi ,\eta ),
\end{equation}
that formally yields
\begin{equation}
\phi_{A}(u,v)=\int \frac{d\xi d\eta }{2\pi }\mathrm{e}^{(\xi
^{2}+\eta ^{2})/4}\tilde{K}_{A}(\xi ,\eta )\mathrm{e}^{i(\xi u+\eta
v)}.
\end{equation}
The presence of the anti-gaussian factor shows that the inverse Fourier
transform of $\tilde{\phi}_{A}(\xi ,\eta )$ exists only when the asymptotic
decay of $\tilde{K}_{A}(\xi ,\eta )$ is faster than the growth of $\mathrm{e}%
^{(\xi ^{2}+\eta ^{2})/4}.$ However, the integral always exists as a
distribution, as was proven in Ref.\cite{MehtaSud65}. By virtue of this
remark, we may go on and substitute the previous expression into Eq.(\ref%
{Defk}) getting
\begin{eqnarray}
&\hat{A}&=\int \frac{d^{2}z}{\pi }\left[ \int \frac{d\xi d\eta }{2\pi }%
\mathrm{e}^{(\xi ^{2}+\eta ^{2})/4}\tilde{K}_{A}(\xi ,\eta )\mathrm{e}%
^{i(\xi u+\eta v)}\right] \left| z\right\rangle \left\langle
z\right| = \\
&&\int \frac{d^{2}z}{\pi } \left[ \int \frac{d\xi d\eta }{2\pi }\int \frac{%
du^{\prime }dv^{\prime }}{2\pi }K_{A}(u^{\prime },v^{\prime
})\mathrm{e}^{(\xi ^{2}+\eta ^{2})/4}\mathrm{e}^{i\left[ \xi \left(u-u^{\prime }\right) +\eta \left( v-v^{\prime }\right) %
\right] }\right] \left| z\right\rangle \left\langle z\right| .  \nonumber
\end{eqnarray}
Upon interchanging the order of integration, we may write the expected
reconstruction formula as
\begin{equation}
\hat{A}=\int \frac{d^{2}z^{\prime }}{\pi }\hat{G}(z^{\prime
})K_{A}(z^{\prime }),  \label{Arec}
\end{equation}
where the operator $\hat{G}(z^{\prime })$ reads:
\begin{equation}
\hat{G}(z^{\prime }):=\int \frac{d^{2}z}{2\pi }\int \frac{d\xi d\eta }{2\pi }%
\mathrm{e}^{(\xi ^{2}+\eta ^{2})/4}\mathrm{e}^{i\left[ \xi \left(u-u^{\prime }\right) +\eta \left( v-v^{\prime }\right) %
\right] }\left| z\right\rangle \left\langle z\right| .  \label{GramCS}
\end{equation}
In other words, the CS tomographic set, like any other tomographic set, is associated with a  resolution of the unity 
\begin{equation}
\mathbb{\hat{I}}=\int \frac{d^{2}z^{\prime }}{\pi }\hat{G}(z^{\prime })%
\mathrm{Tr}(\left| z^{\prime }\right\rangle \left\langle z^{\prime }\right|
\cdot )\quad .
\end{equation}
In a quantizer-dequantizer scheme, the operator $\hat{P}_z=\left| z^{\prime }\right\rangle \left\langle z^{\prime }\right|$ is the dequantizer, while $\hat{G}(z)$ is the quantizer. Since
\begin{equation}
\left\vert z\right\rangle \left\langle z\right\vert =\int \frac{%
d^{2}z^{\prime }}{\pi }\phi_{\left\vert z\right\rangle \left\langle
z\right\vert }\left( z^{\prime }\right) \left\vert z^{\prime }\right\rangle
\left\langle z^{\prime }\right\vert \Leftrightarrow \phi_{\left\vert
z\right\rangle \left\langle z\right\vert }\left( z^{\prime }\right) =\pi
\delta (z-z^{\prime })~,
\end{equation}
we observe that
\begin{eqnarray}
&&K_{{G}(z^{\prime })}(z)=\left\langle z\left\vert \hat{G}(z^{\prime
})\right\vert z\right\rangle =\int \frac{d\xi d\eta }{2\pi }\mathrm{e}^{i%
\left[ -\xi u^{\prime }-\eta v^{\prime }\right] }\mathrm{e}^{(\xi
^{2}+\eta ^{2})/4}\tilde{K}_{\left\vert z\right\rangle \left\langle
z\right\vert }\left( \xi ,\eta \right)  \nonumber \\
&=&\int \frac{d\xi d\eta }{2\pi }\mathrm{e}^{i\left[ -\xi u^{\prime
}-\eta v^{\prime }\right] }\tilde{\phi}_{\left\vert z\right\rangle
\left\langle z\right\vert }\left( \xi ,\eta \right) =\phi_{\left\vert
z\right\rangle \left\langle z\right\vert }\left( z^{\prime }\right) =\pi
\delta (z-z^{\prime })~,
\end{eqnarray}
and remark that this result amounts to the orthonormality relations of the pair quantizer-dequantizer:
\begin{equation}
K_{{G}(z^{\prime })}(z)=\mathrm{Tr}(\left\vert z\right\rangle \left\langle
z\right\vert {\hat{G}}(z^{\prime }))=\pi \delta (z-z^{\prime }).
\end{equation}
If one exchange the roles in the pair, and bears in mind the definition of $\hat{G}(z),$ one recovers the dual symbol of an operator as its Sudarshan's symbol ${\phi}$,
\begin{eqnarray}
&&\mathrm{Tr}({\hat{G}}(z ^{\prime }) \hat{A})=\int \frac{d\xi d\eta }{2\pi }\mathrm{e}^{i%
\left[ -\xi u^{\prime }-\eta v^{\prime }\right] }\mathrm{e}^{(\xi
^{2}+\eta ^{2})/4}\tilde{K}_{A}\left( \xi ,\eta \right)  \nonumber \\
&=&\int \frac{d\xi d\eta }{2\pi }\mathrm{e}^{i\left[ -\xi u^{\prime
}-\eta v^{\prime }\right] }\tilde{\phi}_{A }\left( \xi ,\eta \right) =\phi_{A}\left( z^{\prime }\right) ~.
\label{KQ}
\end{eqnarray}
In this dual theory,  the definition of the Sudarshan's symbol as diagonal representation, Eq.(\ref{Defk}), appears as a reconstruction formula, the quantizer being $\hat{P}_z$.
In a tomographic picture of quantum mechanics, this gives the possibility of representing in tomographic, i.e., "inner"  terms, quantities such as expectation values of observables
\begin{equation}
\mathrm{Tr}(\hat{\rho}\hat{A})=\int \frac{d^{2}z}{\pi }K_{\rho}(z)\mathrm{Tr}(\hat{G}(z)\hat{A})=\int \frac{d^{2}z}{\pi }K_{\rho}(z)\phi_{A}( z) .
\end{equation}
We remark that, in the context of the Agarval-Wolf operator ordering theory  \cite{MandelWolf,AgWolf}, the $K$ and $\phi$ symbols, appearing in our direct and dual reconstruction formulae,
are related to the Wick (i.e., normal) and anti-Wick ordering
respectively.
Of course, if one insists in requiring only the use of CS tomographic representation, one needs to introduce a star-product by means of an integral kernel:
\begin{eqnarray}
&&\mathrm{Tr}(\hat{\rho}\hat{A})=K_{\rho A}(z)=(K_{\rho} \star K_{A})(z):=\int \frac{d^{2}z_1}{\pi }\frac{d^{2}z_2}{\pi }K_{\rho}(z_1)K_{A}(z_2) \mathcal{Q}(z_1, z_ 2,z) \nonumber \\
&&\mathcal{Q}(z_1, z_ 2,z):=\mathrm{Tr}(\hat{G}(z_1)(\hat{G}(z_2)\hat{P}_z) . 
\end{eqnarray}
More details on star-product kernels are contained in a recent work with Volodya Man'ko \cite{Ib13}.
We conclude by observing that the CS tomography can be generalized by using  coherent states and nonlinear coherent states of deformed oscillators, including $q-$oscillators. This generalization was analyzed in \cite{CSTOM}.


\section{Tomography and the algebraic description of quantum systems}\label{sec:algebras}

As it was discussed in the introduction the algebraic description of quantum systems based on the theory of $C^*$-algebras emerges from an analysis of the fundamental structures needed to describe physical systems.    The tomographic description of them has been started recently in \cite{Ib13}, (see also \cite{Ye14} and references therein).    

Jordan algebras were introduced by P. Jordan 
as an attempt to unfold the algebraic structure of quantum systems, but it was only after the work of Alfsen and Schultz \cite{Al98} that the exact relation with the theory of $C^*$-algebras was established.   Recently, such relation was revisited by Falceto et al \cite{Fa13}, \cite{Fa13b} and the relation between $C^*$-algebas and Lie-Jordan Banach algebras was clearly established.   Such equivalence was used too to describe the theory of reduction of $C^*$-algebras described in terms of Lie-Jordan Banach algebras.  

In this section the first steps towards a tomographic description of quantum systems based on $C^*$ and Lie-Jordan algebras will be established.   One of the main outcomes of the theory is that it fits nicely with the tomographic description based on groups as it will be shown in what follows.  Thus, we will review first the basic notions from the theory of $C^*$ and Lie-Jordan algebras that will be needed in what follows and after this the tomographic description of a quantum system described by a Lie-Jordan algebra and a group representation will be sucintly described.

\subsection{$C^*$-tomography}

We will consider a quantum system described by a unital $C^*$-algebra $\mathcal{A}$, whose self--adjoint part constitute the Lie-Jordan-Banach algebra of observables of the theory.  The states of the system are normalized positive functionals $\rho\colon \mathcal{A}\to\mathbb{C}$,
$$
\rho(\mathbb{I})=1 \, , \qquad \rho(a^*a)\geq0 \, , \qquad \forall a\in\mathcal{A}
$$ 
and they determine a weak*--compact subset $S(\mathcal{A})$ of the topological dual $\mathcal{A}'$.

Given $a\in\mathcal{A}$, the number $\rho(a)$ is the expected value of the observable measured in the state $\rho$, also denoted as: $\langle a\rangle_\rho$.  Hence for each self-adjoint element $a\in\mathcal{A}$, we may define a continuous affine function $\hat{a} \colon S(\mathcal{A})\to \mathbb{R}$,  $\hat{a}(\rho) = \rho(a)$.  Kadison theorem \cite{Ka51} states that the correspondence $a\mapsto \hat{a}$ is an isometric isomorphism from the self-adjoint part of $\mathcal{A}$ onto the space of all real continuous affine functions on  $S(\mathcal{A})$.  Notice that the Hilbert space picture of the system, once a state is given, can be recovered by means of the GNS construction \cite{Naimark}.

If we consider now the general background of tomography as stated in Sect. \ref{sec:general}, we may consider that the space of states is a subset of the linear space $V = \mathcal{A}'$.  Moreover because $\mathcal{A} \subset \mathcal{A}''$, then we may think that the elements $\alpha$ in the dual $V^*$ are going to lie in $\mathcal{A} \subset V^*$.

Thus the tomographic description of the state $\rho$ of $\mathcal{A}$ will consist in assigning to this state a probability density function $W_\rho$, that we will call ``tomogram'', on some auxiliary space $\mathcal{N}$ such that given $W_\rho$ the state $\rho$ can be reconstructed unambiguously.

Hence we consider a family of elements in $\mathcal{A}$ parametrized by an index which can be discrete or continuous, the elements of $\mathcal{N}$, or, in other words, consider a map $U\colon \mathcal{N} \to \mathcal{A}$ and we will denote by $U(x)$ the element in $\mathcal{A}$ associated to the element $x\in \mathcal{N}$.  The family $\left\{U(x)\,|\,x\in\mathcal{N}\right\}$ will be called a tomographic set if it separates states, i.e., given $\rho_1,\rho_2\in S(\mathcal{A})$, if $\rho_1\neq\rho_2$, then $\exists\, x\in\mathcal{N}$ such that $\langle\rho_1,U(x)\rangle\neq\langle\rho_2,U(x)\rangle$.

Given a state $\rho$ and a tomographic set $U$, we will call the function $F_\rho \colon \mathcal{N} \rightarrow\mathbb{C}$, defined as, recall Eq. (\ref{general_tomogram}): 
$$
F_\rho(x)=\langle\rho,U(x)\rangle \, ,
$$ 
the sampling function of $\rho$ with respect to $U$.

Let us assume that $\mathcal{N}$ is a topological space and a positive Borel measure $\mu$ on it. We will also assume that the map $U$ is continuous and integrable in the sense that for any $\rho\in S(\mathcal{A})$, the sampling function $F_\rho \colon \mathcal{N} \to \mathbb{C}$ is integrable, that is, $F_\rho\in L^1(\mathcal{N},\mu)$.    The auxiliary space $\mathcal{N}$ (and the measure $\mu$ on it) used in the theory could depend on the specific problem.    Later on a rather general way of selecting the auxiliary spaces by means of groups and their unitary representations will be discussed (see \S \ref{sec:group_unitary}).

In order to guarantee that the assignament $\rho \mapsto F_\rho$ is invertible, we will assume that there exists another map $D\colon \mathcal{N} \to \mathcal{A}'$ which is integrable in the sense that for any $a\in\mathcal{A}$ the function $G_a(x)=\langle D(x),a\rangle$ is integrable, and such that
\begin{equation}
\langle D(x),U(x')\rangle=\delta(x,x'),\quad x,x'\in\mathcal{N} \,,
\end{equation}
where $\delta(x,x')$ is the delta distribution along the diagonal on $\mathcal{N} \times \mathcal{N}$, that is:
\begin{equation}\label{biortho}
\phi(x)=\int_{\mathcal{N}}\delta(x,x')\phi(x')d\mu(x')\,,
\end{equation}
where $\phi$ is a continuous function with compact support on $\mathcal{N}$.
If a map $D$ exists satisfying the property (\ref{biortho}) we will say that $U$ and $D$ are biorthogonal. 

If we denote now by $\check{\rho}(x) = F_\rho(x) = \langle\rho,U(x)\rangle$ and by 
$$
\hat{\phi} = \int_{\mathcal{N}} \phi(x)D(x)d\mu(x)\,,
$$ 
with $\phi$ an integrable function, then it is easy to see that $\check{\hat{\phi}}(x)=\phi(x)$ and $\hat{\check{\rho}}=\rho$. 

It is also noticeable that because 
$$
\int_{\mathcal{N}}F_\rho(x)d\mu(x)=\int_{\mathcal{N}}\langle\rho,U(x)\rangle d\mu(x)=\rho\left(\int_{\mathcal{N}}U(x) d\mu(x)\right) \ ,
$$
it is convenient to assume that 
$$
\int_{\mathcal{N}} U(x)d\mu (x)=\mathds{1}\,.
$$
If $U(x)$ satisfies this, then we will say that $U(x)$ is normalized. In such case it is clear that $$\int_{\mathcal{M}}F_{\rho}(x)d\mu (x)=1\,.$$

We may recast the previous theorem by computing first the $\,\:\:\hat{}\,\:$ map and later the $\,\:\:\check{}\,\:$ map on $F_\rho$:
\begin{equation}
\hat{F}_\rho=\int_{\mathcal{N}}D(x)F_\rho(x)d\mu (x)\,,
\end{equation}
then if we apply the $\,\:\:\check{}\,\:$ map first, we get
\begin{equation}
\hskip-2cm \check{\hat{F}}_\rho(x')=\langle\hat{F}_\rho(x),U(x')\rangle=\int_{\mathcal{N}}F_\rho(x)\langle D(x),U(x')\rangle \, d\mu (x)=F_\rho(x')\,,\quad\forall x'\in\mathcal{N}\,.
\end{equation}

We may also define another function $F_{\rho}(x,x')$ depending of two arguments instead of one:  $F_{\rho}(x,x')=\langle\rho,U(x)^*U(x')\rangle$ for any $x$,$x'\in\mathcal{N}$.  We will say that a function $F\colon \mathcal{N}\times\mathcal{N}\rightarrow\mathds{C}$ is positive, or of positive type, or positive semidefinite, if $\forall N\in\mathbb{N}$ and $\xi_i\in\mathds{C}$, $x_i\in\mathcal{N}$, $i=1\ldots,N$, we have:

\begin{equation}\label{positive}
\sum_{i,j=1}^N\bar{\xi}_i\xi_jF(x_i,x_j)\geq0\,.
\end{equation}

Then it is easy to check that given a state $\rho$ and a tomographic set $U\colon \mathcal{N} \to \mathcal{A}$ in a $C^*$-algebra $\mathcal{A}$, then the function $F_\rho(x,x')=\langle\rho,U(x)^*U(x')\rangle$ is positive semidefinite.
In fact, the following simple computation shows it.
\begin{eqnarray*}
\sum_{i,j=1}^N\bar{\xi}_i\xi_jF_\rho(x_i,x_j) &=& \sum_{i,j=1}^N\bar{\xi}_i\xi_j\langle\rho,U(x_i)^*U(x_j)\rangle=\left\langle\rho,\sum_{i,j=1}^N\bar{\xi}_i\xi_jU(x_i)^*U(x_j)\right\rangle\\
&=& \left\langle\rho,\left(\sum_{i=1}^N\xi_iU(x_i)\right)^*\left(\sum_{j=1}^N\xi_jU(x_j)\right)\right\rangle\geq0\,.
\end{eqnarray*}
We will take advantage of this property later on when dealing with tomography in groups.

\subsection{Equivariant tomographic theories on C*--algebras}

We will encounter in many situations the presence of a group in the theory whose states we want to describe tomographically. Such group could be a group of symmetries of the dynamics of the system or a group which is describing the background of the theory.  In any of these circumstances we will assume that there is a Lie group $G$ acting on the $C^*$-algebra $\mathcal{A}$, i.e., we have a strongly continuous map $T \colon G\to \mathrm{Aut}(\mathcal{A})$ such that 
$$
T_e=\mathbb{I} \, , \qquad T_{g_1}T_{g_2}=T_{g_1g_2}\, , \qquad \forall g_1,g_2\in G \, .
$$  
In such case we will need to assume that the group $G$ acts on the auxiliary spaces used to construct the tomographic picture. Thus, the group $G$ will act on $\mathcal{N}$, and such action will be simply denoted as $x\mapsto g\cdot x$, $x\in\mathcal{N}$.

The natural compatibility condition for a tomographic map $U\colon \mathcal{N}\to \mathcal{A}$ to be compatible with the group $G$ present in the theory is equivariance, i.e.,
\begin{equation}
U(g\cdot x)=T_g(U(x)) \, , \qquad\forall x\in\mathcal{N},\: g\in G\,.
\end{equation}
This could be interpreted by saying that if the parameters $x,x'$ parametrizing two sampling elements $U(x)$ and $U(x')$ in $\mathcal{A}$, are related by an element $g$ of the group, i.e., $x'=g\cdot x$, then the two sampling observables $U(x)$,$U(x')$ are also related by the same element of $G$.

Under these conditions, it is easy to conclude that the sampling function $F_\rho$ corresponding to the state $\rho$ satisfies the following:
\begin{equation}
F_\rho(g\cdot x)=F_{T_g^*\rho}(x)\,,
\end{equation}
because
\begin{equation}
F_\rho(g\cdot x)=\langle\rho, U(g\cdot x)\rangle=\langle\rho,T_g(U(x))\rangle=\langle T_g^*\rho,U(x)\rangle=F_{T_g^*\rho}(x)\,,
\end{equation}
where $T_g^*\rho$ denotes the adjoint action of $G$ on $\mathcal{A}^\prime$. Notice that if $\rho$ is an invariant state, $T_g^*\rho=\rho$ then the corresponding samping function will be invariant too.
\begin{equation}
F_\rho(g\cdot x)=F_\rho(x)\,.\qquad\forall\in G\,,\quad x\in\mathcal{M}\,.
\end{equation}

We will also consider that situation where the group $G$ acts on the space $M$ defining the space of classical functions whose tomograms we would like to obtain.  We will denote again such action by $y\mapsto g\cdot y$, $y\in M$. The generalised Radon transform $\mathcal{R} \colon \mathcal{N} \to \mathcal{F}(M)^\prime$ should be equivariant, this is
\begin{equation}
\mathcal{R}(g\cdot y)=g_*\mathcal{R}(y)\,,
\end{equation}
where $g_*$ indicates now the natural action induced on the space $\mathcal{F}(M)^\prime$ in the dual of the space of functions on $M$, by the action of $G$ on $M$. More explicitly 
\begin{equation}
\langle g_*(\mathcal{R}(y)),F\rangle=\langle\mathcal{R}(y),g^*F\rangle\quad\mbox{and}\quad g^*F(x)=F(g\cdot x)\,.
\end{equation}
Now, if $\mathcal{R}$ is actually a generalized Radon transform and $W_\rho$ denote the tomogram of the state $\rho$, we will have:
\begin{equation}
\hskip-2.3cm W_\rho(g\cdot y)=\langle\mathcal{R}(g\cdot y),F_\rho\rangle=\langle g_*\mathcal{R}(y),F_\rho\rangle=\langle\mathcal{R}(y),g^*F_\rho\rangle=\langle\mathcal{R}(y),F_{g^*\rho}\rangle=W_{g^*\rho}(y)\,.
\end{equation}
Then we conclude this discussion by observing that if $\rho$ is an invariant state, the tomogram defined by it, is actually invariant:
\begin{equation}
W_\rho(g\cdot y)=W_\rho(y)\qquad\forall g\in G\,.
\end{equation}

\subsection{Unitary group representations on $C^*$-algebras}\label{sec:group_unitary}

We will discuss now a particular instance of the tomographic programme discussed above where a Lie group $G$ plays a paramount role. We will find this situation for instance in spin tomography where the group $G$ will be the group $SU(N)$, but such situation is found also in standard homodyne tomography that could be understood in similar terms with the group $G$ being the Heisenberg--Weyl group.

Now we will assume that the auxiliary space $\mathcal{N}$ is a group $G$ and the tomographic map $U \colon G\to \mathcal{A}$ is provided by a continuous unitary representation of $G$ on $\mathcal{A}$, that is, $U(g)^*=U(g)^{-1}=U(g^{-1})$ is a unitary element in the $C^*$-algebra $\mathcal{A}$, and $U(g_1g_2)=U(g_1)U(g_2)$ and $U(e)=\mathds{1}$.  Then we may denote by $T \colon G\to \textrm{Aut}(\mathcal{A})$, the action of $G$ on $\mathcal{A}$ by inner automorphism given by
$$
T_g(a)=U(g)^*\cdot a \cdot U(g) \, ,\qquad  a\in\mathcal{A} \, , g\in G \, .
$$  
Then we can see immediately that we have the equivariance property for the tomographic map with respect to the action of $G$ on itself by conjugation:
\begin{equation}
U(g^{-1}hg)=U(g^{-1})\cdot U(h) \cdot U(g) =T_g(U(h)) \, , \qquad g,h \in G\, .
\end{equation}
The sampling function corresponding to the state $\rho$ is given by
\begin{equation}
F_\rho(g)=\langle\rho,U(g)\rangle\,,
\end{equation}
and we may check that the map $F_\rho:G\rightarrow\mathds{C}$ is a positive semidefinite map in the sense that the map $F_\rho (g, h)$ of two arguments defined as 
$$
F_\rho (g,h) = F_\rho (g^{-1}h) \, \qquad g,h \in G \, , 
$$
is positive semidefinite, i.e. $\forall N\in\mathds{N}$, $\xi_i\in\mathds{C}$, $g_i\in G$, $i=1,\ldots,N$, then
\begin{equation}
\sum_{i,j=1}^N\bar{\xi}_i\xi_jF_\rho(g_i^{-1}g_j)\geq0\,.
\end{equation}

Naimark's theorem \cite{Naimark} establishes that given a positive semidefinite function $F$ on a group $G$, there exists a Hilbert space $\mathcal{H}_F$, a continuous unitary representation $U_F:G\rightarrow\mathcal{U}(\mathcal{H}_F)$ and a vector $|0\rangle\in\mathcal{H}_F$ such that
\begin{equation}
F(g)=\langle0|U_F(g)|0\rangle\,.
\end{equation}
The relation between the state $|0\rangle\langle0| = \rho_F$, the representation $U_F$, the original state $\rho$ and the original representation $U$ has been discussed recently \cite{Ib11}.

In order to define now the generalized Radon transform that will describe the tomogram corresponding to the state $\rho$, we will consider now the Lie algebra $\mathfrak{g}$ of the group $G$ and the auxiliary space $\mathfrak{g}\times\mathds{R}$.  In $\mathfrak{g}\times\mathds{R}$ we have defined the natural extension of the standard exponential map $\widetilde{\exp}:\mathfrak{g}\times\mathds{R}\rightarrow G$ given by $\widetilde\exp(\xi,s)=\exp(s\xi)$.

The unitary representation of $G$ on $\mathcal{A}$, defines a Lie algebra homomorphism from $\mathfrak{g}$ into $\mathcal{A}_{sa}$, where $\mathcal{A}_{sa}$ denotes the self-adjoint part of $\mathcal{A}$, that is, $\mathcal{A}_{sa}=\left\{a\in\mathcal{A}\,|\,a^*=a\right\}$. Given any $\xi\in\mathfrak{g}$, we denote by $\hat{\xi}$ the element of $\mathcal{A}_{sa}$ defined as:
\begin{equation}
\hat{\xi}=-i\left.\frac{\textrm{d}}{\textrm{d}s}U\left(\exp(is\xi)\right)\right|_{s=0} = -i \left.\frac{\partial}{\partial s} U(\widetilde\exp(\xi,s))\right|_{s=0} \,,
\end{equation}

Clearly $\hat{\xi}^* = \hat{\xi}$ and
\begin{equation}
[\hat{\xi},\hat{\zeta}] = \widehat{[\xi,\zeta]}\,,\qquad \forall \xi,\zeta\in\mathfrak{g}\,.
\end{equation}

\subsection{The GNS construction and unitary group representations on $C^*$-algebras}

Given the state $\rho$, the GNS construction allows to represent the $C^*$-algebra $\mathcal{A}$ as operators acting on a Hilbert space $\mathcal{H}_\rho$.  To be precise consider the Hilbert space $\mathcal{H}_\rho$ obtained as the completion of the space $\mathcal{A}/\mathcal{J}_\rho$, where $\mathcal{J}_\rho=\ker\rho$ is the Gelfand ideal defined by $\rho$, with respect to the inner product is defined as:
\begin{equation}
\langle[a],[b]\rangle = \rho(a^*b)\,,\qquad \forall [a]=a+\mathcal{J}_\rho, [b]=b+\mathcal{J}_\rho \in \mathcal{A}/\mathcal{J}_\rho\,.
\end{equation}
Then there is a natural representation $\pi_\rho:\mathcal{A}\rightarrow\mathcal{B}(\mathcal{H})$ given by:
\begin{equation}
\pi_\rho(a)[b] = [a\cdot  b]\,,\qquad\forall[b]\in\mathcal{H}_\rho\,.
\end{equation}
The unitary representation $U \colon G\to \mathrm{Aut}(\mathcal{A})$ of the group $G$ becomes a unitary representation $U_\rho$ of $G$ on $\mathcal{H}_\rho$ by means of $U_\rho=\pi_\rho\circ U$, i.e., $U_\rho:G\rightarrow\mathcal{U(\mathcal{H}_\rho)}$ is the map defined by 
$$
U_\rho(g)=\pi_\rho\left(U(g)\right) \, .
$$ 
Notice that because $U$ is unitary, $U(g)$ is a unitary element in $\mathcal{A}$, then $\pi_\rho\left(U(g)\right)$ is a unitary operator on $\mathcal{H}_\rho$. In fact, for all $[a], [b] \in\mathcal{H}_\rho$ we have:
\begin{eqnarray}
\hskip-1.5cm \langle U_\rho(g)[a],U_\rho(g)[b]\rangle_\rho &=& \langle \pi_\rho\left(U_\rho(g)\right)[a],\pi_\rho\left(U_\rho(g)\right)[b]\rangle_\rho=\langle [U_\rho(g)a],[U_\rho(g)b]\rangle_\rho\\
 &=& \rho\left( \left(U_\rho(g)a\right)^*U_\rho(g)b \right)=\rho(a^*b)=\langle[a],[b]\rangle_\rho .
\end{eqnarray}

If we consider now the induced map $\mathfrak{g} \to \mathcal{A}_{sa}$, we will obtain that the element $\xi \in \mathfrak{g}$ will be mapped by the representation $\pi_\rho$ into a self-adjoint operator on $\mathcal{H}_\rho$.
We will denote the operator defined this way by $\hat{\xi}_\rho$, that is:
\begin{eqnarray}
i\hat{\xi}_\rho\left([a]\right) &=& \left.\frac{\textrm{d}}{\textrm{d}s}U_\rho\left(\exp(s\xi)[a]\right)\right|_{s=0}=\left.\frac{\textrm{d}}{\textrm{d}s}\pi_\rho\left(U\left(\exp(s\xi)[a]\right)\right)\right|_{s=0}\\
&=& \left.\frac{\textrm{d}}{\textrm{d}s}[U\left(\exp(s\xi)\right)a]\right|_{s=0}=\left[\left.\frac{\textrm{d}}{\textrm{d}s}U\left(\exp(s\xi)\right)\right|_{s=0}\cdot a\right]=[i\hat{\xi}a]\,.
\end{eqnarray}

Because of Stone's theorem, we know that there exists a unique strongly continuous one-parameter group of unitary operators $U_s^\xi$ on $\mathcal{H}_\rho$ such that
\begin{equation}
\left.\frac{\textrm{d}}{\textrm{d}s}U_s^\xi[a]\right|_{s=0}=i\hat{\xi}_\rho[a]\,,
\end{equation}
thus:
\begin{equation}
\e^{is\hat{\xi}_\rho} = U_s^\xi = U_\rho\left(\exp(s\xi)\right)=\pi_\rho\circ U\left(\exp(s\xi)\right)\,.
\end{equation}

Finally, the spectral theorem applied to the self-adjoint operators $\hat{\xi}_\rho$ asserts that there is a Borelian spectral measure $E_{\xi_\rho}$ in the real line such that
\begin{equation}
\hat{\xi}_\rho=\int \, \, \lambda \, \mathrm{d}E_{\xi_\rho}(\lambda)\,,
\end{equation}
hence
\begin{equation*}
U_s^\xi=\int\e^{is\lambda}\textrm{d}E_{\xi_\rho}(\lambda)\,.
\end{equation*}
Now, we will consider as auxiliary space $\mathcal{N}$ the space $\mathfrak{g}\times\mathbb{R}$, and we will define the map $\mathcal{R} \colon \mathfrak{g}\times\mathbb{R}\rightarrow \left(\mathcal{F}(G)\right)^\prime$ as follows:
\begin{equation}
\mathcal{R}(\xi;\lambda)(F)=\frac{1}{(2\pi)^2}\int\e^{-is\lambda}F \left(\exp(s\xi)\right)\textrm{d}s\,.
\end{equation}
To be precise, the map $\mathcal{R}$ is defined from $\mathfrak{g}\times\mathds{R}$ into the space of continuous functions on the exponential of $\mathfrak{g}$, $\exp(\mathfrak{g})\subset G$. For groups such that $\exp(\mathfrak{g})=G$, i.e. exponential groups, we have the map written above.   

Then we will call the function $\mathcal{R}(\xi;\lambda)(F_\rho)$ corresponding to the tomographic function $F_\rho$ defined by the state $\rho$, the tomogram of $\rho$ and will be denoted by $W_\rho (\xi; \lambda)$.
Notice that we may fix $\lambda=1$ to get:
\begin{equation}
W_\rho(\xi)=\frac{1}{(2\pi)^2}\int\e^{-is}F_\rho\left(\exp(is\xi)\right)\textrm{d}s.
\end{equation}

Notice that if we compute the inverse Fourier transform of the tomogram $W_\rho$, we have:
\begin{equation}
F_\rho\left(\exp(s\xi)\right)= \int\e^{is\lambda}W_\rho(\xi;\lambda)\, \mathrm{d}\lambda\,
\end{equation}
and if we consider now that
\begin{eqnarray}
F_\rho\left(\exp(s\xi)\right) &=& \langle\rho,U\left(\exp(s\xi)\right)\rangle=\langle0|U_\rho\left(\exp(s\xi)\right)|0\rangle\\
&=& \langle0|\int\e^{is\lambda}\textrm{d}E_{\xi_\rho}(\lambda)|0\rangle=\int\e^{is\lambda}\langle0|\textrm{d}E_{\xi_{\rho}}(\lambda)|0\rangle\,.
\end{eqnarray}

Then if the measure $\mu_{\xi,\rho}=\langle0|\textrm{d}E_{\xi_{\rho}}|0\rangle$ is absolutely continuous with respect to the Lebesgue measure, we will have that $\mu_{\xi,\rho}(\lambda)=W_\rho(\xi;\lambda)\textrm{d}\lambda$, and  $W_{\rho}(\xi;\lambda)$ will be the Radon-Nikodym derivative of $\mu_{\xi,\rho}$ with respect to $\textrm{d}\lambda$. Hence, we have obtained that, under these conditions, the tomogram $W_\rho$ of the state $\rho$ can be written as:
\begin{equation}
W_\rho(\xi;\lambda)=\frac{\delta \mu_{\xi,\rho}}{\delta\lambda}\,,
\end{equation}
where $\delta \mu_{\xi,\rho}/\delta\lambda$ denotes the Radon-Nikodym derivative of $\mu_{\xi,\rho}$ with respect to the Lebesgue measure $\textrm{d}\lambda$.

Finally, suppose that $\omega$ is a density operator on $\mathcal{H}_\rho$.   We will define its tomographic function by $F_\omega (g) = \mathrm{Tr} (\omega U_\rho (g))$ and its tomogram $W_\omega (\xi; \lambda)$ as;
$$
W_\omega (\xi; \lambda) = \frac{1}{(2\pi)^2}\int\e^{-is\lambda}F_\omega \left(\exp(s\xi)\right)\textrm{d}s\,.
$$
Then computing the inverse Fourier transform of $W_\omega (\xi; \lambda)$ we will get:
$$
F_\omega \left(\exp(s\xi)\right)= \int\e^{is\lambda}W_\omega (\xi;\lambda)\, \mathrm{d}\lambda\, .
$$
A simple computation shows that:
\begin{eqnarray*}
\hskip-1.5cm W_\omega (\xi; \lambda) &=& \frac{1}{(2\pi)^2}\int \e^{-is\lambda}F_\omega \left(\exp(s\xi)\right)\textrm{d}s 
=  \frac{1}{(2\pi)^2}\int\e^{-is\lambda}  \mathrm{Tr} \left(\omega U_\rho\left(\exp(s\xi)\right) \right) \textrm{d}s  \\ 
&=& \mathrm{Tr} \left( \frac{1}{(2\pi)^2}\int\e^{-is\lambda} \omega U_\rho\left(\exp(s\xi)\right) \textrm{d}s \right) 
= \mathrm{Tr} \left( \frac{1}{(2\pi)^2}\int \e^{-is\lambda}  \omega \e^{is\hat{\xi}_\rho} \textrm{d}s \right) \\
&=&  \mathrm{Tr} \left( \omega  \frac{1}{(2\pi)^2}\int\e^{-is(\lambda\mathbb{I} - \hat{\xi}_\rho)} \textrm{d}s \right) 
=  \mathrm{Tr} \left( \omega \, \delta (\lambda \mathbb{I} - \hat{\xi}_\rho) \right)  \, ,
\end{eqnarray*}
which is the formula for the tomogram $W_\rho$ that resembles closely the classical Radon transform Eq. (\ref{eq:radondef}) in the affine language.

\ack{We  thank Volodya and Margarita Man'ko for many invaluable discussions and  intense collaboration
on tomography and related subjects. This work was partially supported by MEC grant
MTM2010-21186-C02-02 and QUITEMAD+ programme.  The work of MA  has been partially supported by  Spanish DGIID-DGA grant 2014-E24/2 and  Spanish
MICINN grants FPA2012-35453 and CPAN-CSD2007-00042.}


\end{document}